\ifx\onecol\undefined
\documentclass[journal,twocolumn]{IEEEtran}
\else 
\documentclass[12pt,draftclsnofoot,journal,onecolumn]{IEEEtran}
\fi

\IEEEoverridecommandlockouts

\usepackage{amsmath}
\usepackage{amsthm,amssymb}
\usepackage{algorithm}
\usepackage{algorithmic}
\usepackage{bm,bbm}
\usepackage{balance}
\usepackage{braket}
\usepackage{booktabs}
\usepackage{color}
\usepackage{cases}
\usepackage{eqparbox}
\usepackage{graphicx}
\usepackage{mathrsfs}
\usepackage{multirow}
\usepackage{subfigure}
\usepackage{setspace}
\usepackage{threeparttable}
\usepackage{verbatim}

\usepackage[numbers,sort&compress]{natbib}

\newtheorem{lemma}{Lemma}

\newtheorem{remark}{Remark}

\usepackage{hyperref}
\hypersetup{hidelinks, 
colorlinks=true,
allcolors=black,
pdfstartview=Fit,
breaklinks=true}

\ifx\onecol\undefined
    \newcommand{\widthRatio}{1}
\else
    \newcommand{\widthRatio}{0.7}
\fi
\newcommand{\ri}{{\mathrm{i}}}

\newcommand{\red}[1]{{#1}}

\newcommand{\diff}{{\mathrm{d}}}
\newcommand{\tr}{{\rm{tr}}}

\def \T {^{\mathsf{T}}}
\def \H {^{\mathsf{H}}}

\begin{document}

\title{Electromagnetic Information Theory-Based Statistical Channel Model for Improved Channel Estimation }

\author{{Jieao~Zhu, Zhongzhichao~Wan,~Linglong~Dai,~{\textit{Fellow,~IEEE}}, and~Tie~Jun~Cui,~{\textit{Fellow,~IEEE}} }
\thanks{
    The conference version of this work was presented at the IEEE ICC'24, Denver, CO, USA, 9-13 June 2024~\cite{zhu2024EITGPR}. 
}
\thanks{This work was supported in part by the National Key Research and Development Program of China (Grant No.2023YFB3811503), in part by the National Natural Science Foundation of China (Grant No. 62031019), and in part by the National Natural Science Foundation of China under Grant 62288101. }
\thanks{J. Zhu, Z. Wan, and L. Dai are with the Department of Electronic Engineering, Tsinghua University, Beijing 100084, China, and also with the Beijing National Research Center for Information Science and Technology (BNRist), Beijing 100084, China (e-mails: zja21@mails.tsinghua.edu.cn, wzzc20@mails.tsinghua.edu.cn, daill@tsinghua.edu.cn).}
\thanks{T. J. Cui is with the State Key Laboratory of Millimeter Waves, Southeast University, China (e-mail: tjcui@seu.edu.cn).}
}

\maketitle

\begin{abstract}
    Electromagnetic information theory (EIT) is an emerging interdisciplinary subject that integrates classical Maxwell electromagnetics and Shannon information theory. The goal of EIT is to uncover the information transmission mechanisms from an electromagnetic (EM) perspective in wireless systems. Existing works on EIT are mainly focused on the analysis of EM channel characteristics, degrees-of-freedom, and system capacity. However, these works do not clarify how to integrate EIT knowledge into the design and optimization of wireless systems. To fill in this gap, in this paper, we propose an EIT-based statistical channel model with simplified parameterization. Thanks to the simplified closed-form expression of the EMCF, it can be readily applied to various channel modeling and inference tasks. Specifically, by averaging the solutions of Maxwell's equations over a tunable von Mises distribution, we obtain a spatio-temporal correlation function (STCF) model of the EM channel, which we name as the EMCF. Furthermore, by tuning the parameters of the EMCF, we propose an EIT-based covariance estimator (EIT-Cov) to accurately capture the channel covariance. Since classical MMSE estimators can exploit prior information contained in the channel covariance matrix, we further propose the EIT-MMSE channel estimator by substituting EMCF for the covariance matrix. Simulation results show that both the proposed EIT-Cov covariance estimator and the EIT-MMSE channel estimator outperform their baseline algorithms, thus proving that EIT is beneficial to wireless communication systems. 
\end{abstract}

\section{Introduction}

\red{Massive multiple-input multiple-output (mMIMO) technology has become a reality in existing 5G~\cite{andrews2014will} and is anticipated to continue playing its key role in the future 6G~\cite{jiang2021road}. Thanks to the large number of Tx/Rx antennas, mMIMO systems have provably achieved higher transmission rate and higher reliability by jointly utilizing multiple spatial degrees-of-freedom (DoF)~\cite{lu2014overview}. To fully unleash the spatial DoF offered by the wireless channel, many researchers have recently conceived a new concept called continuous-aperture MIMO (CAP-MIMO)~\cite{pizzo2022fourier,sayeed2010continuous,demir2022channel,huang2020holographic}, where the number of antennas is allowed to grow indefinitely within a finitely large Tx/Rx aperture. This limit regime of mMIMO is made realistic by the recently expanding metasurface antenna technologies~\cite{hunt2014metamaterial}, where massive sub-wavelength radiating structures, such as reconfigurable intelligent surfaces (RISs)~\cite{pei2021ris} and leaky waveguide arrays~\cite{hwang2020binary}, are employed to achieve refined electromagnetic manipulation resolution~\cite{badawe2016true} in a finitely constrained aperture. Due to the complicated interaction~\cite{kelley1993array} between massive closely-spaced sub-wavelength metasurface antennas and the impinging information-carrying EM waves, it is more appealing to model the EM channel responses as continuously distributed fields rather than traditional discrete MIMO matrices, particularly when the arrays are dense~\cite{bjornson2024towards}. This leads to the recently proposed EM-compliant channel modeling methodologies~\cite{pizzo2020spatially,pizzo2022fourier,pizzo2022spatial}, where traditional parametric (ray-tracing)~\cite{fuschini2008analysis} and non-parametric (spatially correlated)~\cite{forenza2007simplified} channel models are re-visited and re-written in an continuously EM-compliant way. This shift in channel modeling methodology happens to coincide with the continuous-time modeling philosophy of information theory in its very beginning~\cite{wyner1966capacity,ihara1993information}, where message-carrying signals are analyzed in the continuous-time domain under the constraint of physical bandlimitation~\cite{slepian1961prolate}. The research of continuously EM-modelled information-theoretic analysis is currently known as electromagnetic information theory (EIT)~\cite{gruber2008new,zhu2022electromagnetic,wang2024electromagnetic,migliore2020cares} or electromagnetic signal and information theory (ESIT)~\cite{di2024electromagnetic}, where fundamental Shannon information-theoretic limits for wireless systems are evaluated subject to Maxwellian constraints.}

\subsection{Prior Works}
\red{EIT research can be divided into two categories: EM channel analysis, and EIT DoF/capacity analysis. }

{\bf EM channel analysis. } Accurate EM-compliant channel models are prerequisites for subsequent information-theoretic analysis. To better capture the EM properties of the wireless channel, novel EM viewpoints of the channels have been intensively researched. Unlike traditional viewpoints that treat the channel as discrete spatio-temporal impulse responses or complex-valued matrices, in~\cite{li2023electromagnetic}, the channel is viewed as a deterministic continuous EM field to describe the scattering properties of the EM MIMO channel. The EM field is expanded by Hankel functions in cylindrical coordinates, and then a $T$-matrix analysis is performed to accurately compute the information-theoretic quantities associated with the MIMO channel. 
To further take into account random EM spatial correlation properties, the authors of~\cite{pizzo2020spatially} provide a novel Gaussian process (GP) viewpoint of wireless channels. This GP representation is derived from the first principles of Maxwell's equations, ensuring its physical consistency. 
In~\cite{pizzo2022fourier}, the authors extend the GP representation in~\cite{pizzo2020spatially} by numerically discretizing the continuous wavenumber domain, and transforming the continuous EM channel into the Fourier plane-wave representation. To complete the Fourier plane-wave model in~\cite{pizzo2022fourier}, various non-ideal effects, including mutual coupling and correlated fading, are explored in recent works~\cite{wang2022electromagnetic,castellanos2023electromagnetic}. 
Another viewpoint is to treat the EM channel as a linear operator, or more precisely, a circle-shaped space-invariant (LSI) filter~\cite{pizzo2022spatial,franceschetti2017wave}. In the wavenumber domain, the EM propagation is characterized by an LSI with asymptotically circular bandpass property, which simplifies the subsequent EIT analysis by connecting EM channel analysis with classical linear filter theory.

{\bf EIT DoF and capacity analysis.} \red{The DoF analysis of physically consistent channels belongs to classical information theory that even dates back to 1946~\cite{gabor1946theory} in the pre-Shannon era. After Shannon's seminal work in 1948 that established the fundamental coding theorems for information theory, the DoF problem, i.e., evaluating the number of independent scalars that can be transmitted through the channel per unit temporal/spatial resource, becomes increasingly important. In the renowned tetralogy~\cite{slepian1961prolate,landau1961prolate,landau1962prolate3,slepian1964prolate4}, D. Slepian, H. O. Pollak and H. J. Landau studied the bandlimited property of temporal waveform channels, and perfectly addressed the associated information-theoretic DoF problem. These works established the standard techniques in dealing with channels with physical bandwidth constraints, and they triggered the subsequent spatial bandwidth analyses for communication systems by considering more practical EM constraints~\cite{bucci1987spatial,franceschetti2015landau,franceschetti2017wave}. 
The first study of EIT DoF analysis is attributed to O. Bucci and G. Franceschetti in 1987~\cite{bucci1987spatial}, where a general scattered EM field observed on a manifold external to a source sphere was analyzed by a bandlimited approximation analysis. The bandwidth of the approximation bases is termed ``spatial bandwidth''. }
Based on the asymptotic analysis in~\cite{bucci1987spatial}, the authors of~\cite{bucci1989degrees} further provided non-asymptotic analysis to the DoF of a general scattered field, given that the spatial bandwidth was known in advance. 
Both of the above works are aimed at evaluating the observed DoF, i.e., the essential dimension of the space which contains all the possible observed fields. Since no constraints were imposed on the source of the field, the transmitter had not been considered. 
\red{To address this problem, a general EIT analysis framework based on solving eigen-systems of the transceiver has been proposed in~\cite{miller2000communicating, gruber2008new, franceschetti2017wave}, and was generalized to arbitrarily shaped transceiver surfaces in~\cite{mikki2023shannon}. 
The instantaneous DoF and capacity are calculated by considering the wireless channel as a spatio-temporal linear operator, and performing an eigen-analysis on this operator. Following this theoretical foundation, the authors of~\cite{jensen2008capacity} proposed a numerical technique for solving electromagnetic eigen-systems to determine the capacity between transceivers. Further extensions to ergodic capacity analyses with randomized EM-compliant channel models are studied in~\cite{poon2005degrees, poon2006impact, nam2014capacity}. }

In summary, existing works on EIT are mainly focused on EM-compliant channel characterization and information-theoretic analysis of wireless systems. These works provide us with new insights and understanding from the physical perspective. Besides the theoretical progress above, people are also very interested in how to exploit EIT for practical system performance improvement. 
\red{Unfortunately, up to now, there are no existing works that integrates EIT knowledge into the design and optimization of wireless systems. This gap is caused by the intractability of acquiring all the required EIT model parameters. 
Despite the accuracy of existing EM-compliant channel models, they depend on precise knowledge of the complicated EM boundary conditions~\cite{mikki2023shannon,miller2000communicating,wang2022electromagnetic,castellanos2023electromagnetic,li2023electromagnetic}, which cannot be perfectly acquired in practice. }

\subsection{Our Contributions}
\red{To fill in this gap, in this paper, we propose an EIT-based statistical channel model with simplified parameterization without sacrificing the accuracy, and demonstrate its beneficial effect in channel estimation}\footnote{Simulation codes will be provided to reproduce the results in this paper: \url{http://oa.ee.tsinghua.edu.cn/dailinglong/publications/publications.html}.}. The contributions of this paper are summarized as follows. 

\begin{itemize}
    \item Unlike previous heuristic constructions of the channel spatio-temporal correlation functions (STCF), we construct a novel channel STCF by averaging the eigen solutions of Maxwell's equations over a tunable probability distribution, which we term as the {\it EM correlation function (EMCF)} of the EM channel. Since the EMCF is directly derived from EM physics, it can accurately model more fundamental propagation characteristics of wireless channels, including tri-polarization and angular sparsity. Thus, the proposed EMCF better matches the actual wireless propagation environment.
    \item \red{By optimizing the tunable parameters of the EMCF, the EMCF channel model can be better fitted to different propagation environments. Specifically, we propose an EIT-based covariance estimator (EIT-Cov) that fits the EMCF to the observed noisy channel data by likelihood maximization. The proposed EIT-Cov integrates EIT knowledge into the channel covariance estimates, thus producing better stochastic description of the EM channel. } 
    \item Since classical MMSE estimators require channel covariance matrices as prior information, we replace this covariance matrix with the proposed EMCF to integrate EM prior information into the MMSE estimators. Based on this replacement, we propose the EIT-MMSE channel estimator to fully exploit the side information contained in the EMCF, thus introducing EM information in a physically interpretable manner. 
    \item \red{Performance analysis and numerical experiments verify that, the proposed EIT-Cov covariance estimator outperforms the baseline covariance estimators, and that the proposed EIT-MMSE channel estimator outperforms the traditional MMSE estimator and compressed sensing estimators.}  It is demonstrated that EIT can be efficiently utilized to improve the performance of wireless communication systems. 
\end{itemize}

\subsection{Organization and Notation}
\emph{Organization}:
The rest of the paper is organized as follows.
\red{
In Section~\ref{sec2-System-Models-and-Problem-Formulation}, we introduce the general construction principle of EIT channel models, and formulate the uplink channel estimation problem. 
In Section~\ref{sec3-the-Proposed-EM-Kernel}, we first introduce basic mathematical concepts of Gaussian random fields (GRF) and their kernel representation. Then, the EMCF is constructed, together with its extension to fast fading channels. Analytical properties and generality of the EMCF is presented. 
In Section~\ref{sec4-Proposed-EIT-GPR-Channel-Estimation-Algorithm}, we propose the EIT-Cov covariance estimator and the EIT-MMSE channel estimator with the EMCF parameter tuning algorithm. 
In Section~\ref{sec5-Simulation-Results}, simulation results are provided for quantifying the performance of the proposed EIT-Cov covariance estimator and the proposed EIT-MMSE channel estimator. 
Finally, in Section~\ref{sec6-conclusion}, we conclude this paper with promising future research ideas. } 

\emph{Notation}: Bold uppercase characters ${\bf X}$ denote matrices;
bold lowercase characters ${\bf x}$ denote vectors; ${\bf X}\T$, ${\bf X}\H$, and ${\bf X}^*$ denotes the transpose, conjugate transpose, and conjugate of matrix ${\bf X}$; 
${\bf x}(n)$ represents the $n$-th component of the vector ${\bf x}$; 
$\Re\{\cdot\}$ and $\Im\{\cdot\}$ denote the real and imaginary part of their arguments, respectively; 
The dot $\cdot$ denotes the scalar product of two vectors; 
$\partial_k$ is the abbreviation of $\partial/\partial x^k$, where $(x^1, x^2, \cdots, x^k, \cdots, x^n)$ is a coordinate vector. 
${\mathbb E}\left[X\right]$ denotes the mean of random variable $X(\omega)$; 
$c$ is the speed of light in a vacuum; 
$[N]$ denotes the positive integer set $\{1, 2, \cdots, N\}$; 
$\mathscr{F}[f(x)]$ denotes the Fourier transform of $f(x)$; 
For spatial distributions, the Fourier basis is $e^{\ri {\bf k}\cdot{\bf x}}$, and for temporal signals, the Fourier basis is $e^{-\ri \omega t}$; 
For ${\bm{x}}\in\mathbb{C}^n$ or $\mathbb{R}^n$, $|{\bm{x}}|,$ denotes the pseudo-norm $\sqrt{{\bm{x}}\T{\bm{x}}}\in\mathbb{C}$, $\left\|{\bm{x}}\right\|$ denotes the standard vector 2-norm $\sqrt{{\bm{x}}\H{\bm{x}}} \in \mathbb{R}_{\geq 0}$; 
$\hat{\bf x}$ represents ${\bf x}/|{\bf x}|$ unless otherwise stated;
For a set $A$, $|A|$ denotes its cardinality; 
$\nabla$ is the gradient operator, and $\nabla \times$ is the curl operator; 
$J_\nu(x)$ is the $\nu$th-order Bessel function of the first kind; 
$\mathbb{S}_{+}^n$ is the collection of all the complex positive semi-definite definite matrices. 
$S^n$ denotes the $n$-dimensional unit sphere embedded in $\mathbb{R}^{n+1}$.

\section{System Models and Problem Formulation} \label{sec2-System-Models-and-Problem-Formulation}
In this section, we will first review the principles of EIT channel analysis, where two different kinds of channel models and their distinct roles in designing channel estimation algorithms are summarized. Then, the uplink channel estimation problem is formulated. 

\subsection{Channel Models} 
Channel models can be divided into two main categories according to their applications. The first kind is the simulation channel model (SCM), which accurately models the EM propagation channel for simulation and industrial evaluation purposes~\cite{CDL,molisch2006cost259,jaeckel2014quadriga}, among which the most representative example is the TR 38.901 model~\cite{CDL}. However, this kind of SCM model is usually over-parameterized, i.e., the number of model parameters is much larger than the number of measurable channel coefficients. Due to large numbers of redundant parameters, SCMs cannot be directly applied to designing channel estimation algorithms. 

To solve the complexity curse of SCMs, simpler models~\cite{kermoal2002stochastic,saleh1987statistical} are usually adopted to enable computationally feasible channel estimation, where one of the most often applied model is the Saleh-Valenzuila channel model (SV model,~\cite{saleh1987statistical}). We name these simpler models as computational channel models (CCMs). CCMs strike a balance between model accuracy and model complexity. Thus, properly constructing CCMs is of practical importance in designing efficient channel estimation algorithms.

An important question is the evaluation of a CCM, i.e., how to assure that a CCM is accurate and satisfactory. To answer this question, let us denote the generative SCM as $p({\bf h})$, where ${\bf h}$ is some channel vector\footnote{We will not dive deep into the definition here, since this probabilistic representation $p({\bf h})$ is general enough to denote any type of channel distributions.}, and the probability measure $p$ assigns a probability value (or density) to each channel realization ${\bf h}$. The true distribution $p$ is usually unknown due to the complexity and dynamicity of the wireless propagation environment. This absence of the true SCM probability $p$ motivates the construction of an approximated CCM probability $q$, which can be found in principle via
\begin{equation}
    q^\star = \arg\min_{q} \left(\lambda D_{\rm KL}(p||q) + (1-\lambda) H(q)\right), \, \lambda\in [0, 1],
    \label{eqn:optimal-CCM}
\end{equation}
where $D_{\rm KL}(\cdot || \cdot)$ denotes the KL divergence between two distributions~\cite{van2014renyi}, and $H(\cdot)$ is the entropy functional. In the above equation~\eqref{eqn:optimal-CCM}, the first term $D_{\rm KL}(p||q)$ corresponds to the accuracy of the surrogate model $q$, and the second term $H(q)$ measures the model complexity. The parameter $\lambda\in [0, 1]$ achieves a trade-off between model accuracy and complexity. 

\red{In this paper, we aim to construct an electromagnetic information theory (EIT)-based CCM $q(\cdot)$. Compared to existing channel models, the proposed model is explainable in the framework of EM theory, and it will exhibit better channel inference performance due to the {\it EM side information}~\cite{lapidoth2002fading} provided by $q$. }

\subsection{Uplink Channel Estimation}
\red{To demonstrate the effectiveness of the proposed EIT-based CCM, we will apply it to the uplink channel estimation problem in a narrowband system. Assume an $N_{\rm BS}$-antenna base station (BS), where each antenna is connected to a dedicated radio-frequency (RF) chain. The uplink signal model is 
\begin{equation}
    {\bf y} = {\bf h} + {\bf n}, 
    \label{eqn:uplink-channel-model}
\end{equation}
where ${\bf y}\in\mathbb{C}^{N_{\rm BS}\times 1}$ is the BS received pilots, ${\bf h}\in\mathbb{C}^{N_{\rm BS}\times 1}$ is circularly symmetric Gaussian distributed with covariance ${\bf R}_{\bf h}\in\mathbb{S}_+^{N_{\rm BS}}$, and ${\bf n}$ is the circularly symmetric complex additive white Gaussian noise (AWGN) with covariance $\gamma^{-1} {\bf I}_{N_{\rm BS}}$, which is independent of the channel ${\bf h}$. The symbol $\gamma$ represents the BS's receive signal-to-noise ratio (SNR) during uplink channel estimation. }

\red{
The goal of uplink channel estimation is to construct a channel estimator $\hat{\bf h}$ from the received pilot signal ${\bf y}$. Given the prior covariance information ${\bf R}_{\bf h}$, the Bayesian optimal channel estimator is the minimum mean-square error (MMSE) estimator, which is given by 
\begin{equation}
    \hat{\bf h}^{\rm MMSE} := {\bf R}_{\bf h}\left({\bf R}_{\bf h} + \frac{1}{\gamma} {\bf I}\right)^{-1}{\bf y}. 
    \label{eqn:standard-MMSE-estimator}
\end{equation}
}
In practical systems, the channel covariance ${\bf R}_{\bf h}$ depends on multiple system parameters, e.g., operating frequency, user mobility, positions of the scatterers, array geometry, and antenna mutual coupling. Although most of these factors also vary with time, the speed of change is usually dozens to hundreds of times slower than that of the instantaneous channel ${\bf h}$~\cite{hesketh2012adaptive}. Thus, it is beneficial to exploit dozens of historical channel realizations for covariance estimation within one statistical coherence block. Note that typical covariance estimators, such as the sample covariance method, need dozens of samples to work. This renders the problem that, after the receiver completes its estimation, the true channel statistics would possibly have changed to another value. There are two solutions to this untimely covariance estimation problem: channel covariance prediction, and advanced channel covariance estimation. The covariance prediction method is a subset of channel predictors that rely on some model of how the channels change over time. \red{The advanced channel covariance estimation method, however, aims at obtaining the true covariance matrix from very few channel observations with the help of some prior information on the channel structure, e.g., narrowband sparsity structure in the angle domain~\cite{haghighatshoar2018low}, or wideband sparsity structure in the angle-delay domain~\cite{haghighatshoar2017massive}. In this paper, we deal with the covariance estimation problem with the help of EIT channel prior information, where the simulation results are compared to the traditional sample covariance method as well as the forward-backward splitting (FBS)-based covariance estimator proposed in~\cite{haghighatshoar2017massive}. }


\section{The Proposed EMCF} \label{sec3-the-Proposed-EM-Kernel}
To construct the EIT prior knowledge, in this section, we will first introduce the EIT channel modeling method based on circularly symmetric Gaussian random fields (CSGRFs). It is proven in Section~\ref{sec:3-1} that CSGRFs are uniquely determined by their kernels. Based on this observation, in Subsection~\ref{sec:3-2}, we propose a novel EMCF (or EM kernel) that captures the EM propagation characteristics. The properties of the proposed EMCF are analyzed in detail in Subsection~\ref{sec:3-3}.

\subsection{Gaussian Random Fields and the Kernel Method} \label{sec:3-1}
\red{Random fields, which have been strongly suggested in~\cite{mikki2023shannon} as the fundamental mathematical language for EIT, are mathematical objects that can naturally characterize the random properties of wireless channels, independent of the antenna geometry~\cite{adler2007random,pizzo2020spatially}. 
Let $h({\bf x},\omega):\, D\times\Omega\to \mathbb{C}$ be a circularly symmetric Gaussian random field (CSGRF) defined on a compact set $D\subset \mathbb{R}^n$, 
where $\Omega$ is the sample space. Specifically, if we arbitrarily pick $K$ points $\{{\bf x}_k\}_{k=1}^K\subset D$, then the joint distribution of the $K$ CSGRF values $\left(h({\bf x}_1), h({\bf x}_2), \cdots, h({\bf x}_K)\right)$ is  multi-variate Gaussian}\footnote{GRFs are equivalent with multi-dimensional Gaussian processes.} distributed: $h({\bf x})\sim \mathcal{GRF}(0, k({\bf x, x'}))$. Similar to multi-variate Gaussian distributions, the probability measure of GRFs are completely determined by their autocorrelation function 
\begin{equation}
    R_h({\bf x}; {\bf x'}) := \mathbb{E}\left[h({\bf x})h^*({\bf x'})\right].
    \label{eqn:mean-covariance-functions}
\end{equation}
The autocorrelation function $R_h: D\times D \to \mathbb{C}$ is usually called the {\it kernel} of the CSGRF $h$. Note that not all of the bi-variate functions are kernels of some CSGRF. To represent some CSGRF, the kernel function must be positive semi-definite~\cite{williams2006gaussian}.  

In order to model the wireless channel with CSGRFs, some restrictions must be imposed on the CSGRF kernel to make the generated random functions $h({\bf x}, t, \omega)$ subject to the EM propagation constraints. Suppose we want to model the electric field distribution by a CSGRF $h({\bf x}, t)$, where $({\bf x}, t)\in\mathbb{R}^4$ is a four-dimensional vector\footnote{This four-dimensional vector is called the spacetime coordinate in Einstein's special and general relativity.}. \red{It can be derived from Maxwell's equations that, a time-harmonic electric field ${\bf E}({\bf x}, t) = {\bf E}(x_1, x_2, x_3, t): \mathbb{R}^4 \to \mathbb{C}^3$ with fixed frequency $f_0$ satisfies the Helmholtz equation
\begin{equation}
    \nabla^2 {\bf E} + k_0^2 {\bf E} = {\bf 0}, \label{eqn:Helmholtz_equation}
\end{equation}
where $k_0 = 2\pi f_0/c$ is the wavenumber. By applying the Helmholtz operator to the random field $h({\bf x}, \omega)$ for each $\omega$ and taking the probabilistic average, we obtain from~\eqref{eqn:mean-covariance-functions} that 
\begin{equation}
    (\nabla^2 + k_0^2)R_h({\bf x}; {\bf x'}) = 0.
    \label{eqn:correlation-Helmholtz}
\end{equation}
}

\begin{remark}
    The necessary condition of a given autocorrelation function to be EM-compliant is Eqn.~\eqref{eqn:correlation-Helmholtz}. This condition is also stated in~\cite{pizzo2022fourier}. 
\end{remark}

Furthermore, if the CSGRF is vector-valued, then its autocorrelation is defined as ${\bf R}_{\bf E}({\bf x}; {\bf x'}) = \mathbb{E}\left[{\bf E}({\bf x}){\bf E}({\bf x'})\H\right]$. In this definition, the autocorrelation function is upgraded to an autocorrelation matrix ${\bf R}_{\bf E}\in\mathbb{C}^{3\times 3}$, where the matrix element $({\bf R}_{\bf E})_{ij}, \,i, j\in \{1, 2, 3\}$ or $i, j\in \{x, y, z\}$ represents the cross-polarized correlation between polarization $i$ and $j$. Note that by definition we have ${\bf R}_{\bf E}({\bf x}, {\bf x}') = {\bf R}_{\bf E}\H({\bf x}', {\bf x})$. Although this matrix-valued autocorrelation seems cumbersome, it is possible to construct greatly simplified mathematical expressions by making physically reasonable assumptions, which will be presented in the next subsection. 

\red{
\begin{remark}
    Another necessary condition of a given matrix-valued autocorrelation function ${\bf R}_{\bf E}$ to be EM-compliant can be derived from the Gauss's law in Maxwell equations to be
    \begin{equation}
        \nabla\cdot {\bf R}_{\bf E}({\bf x}, {\bf x}') = {\bf 0}, \quad\forall ({\bf x}, {\bf x'})\in D^2,  
    \end{equation}
    where $\nabla\cdot$ is the divergence operator w.r.t. variable ${\bf x}$. 
\end{remark}}

\subsection{Construction of the EMCF} \label{sec:3-2} 
To meet the requirements for an EM-compliant general-purpose CCM, we begin by constructing a novel CCM based on the above matrix-valued GRF theory. 
The proposed channel model should be both deeply rooted in EM theory and highly applicable to existing channel estimation algorithms. Thanks to the GRF framework, a feasible approach is to specify the autocorrelation matrix ${\bf R}$ of the ``channel field''. In this way, we can encode the EM knowledge into the autocorrelation function of the channel response. From a signal processing viewpoint, this is equivalent to specifying the correlation matrix of the channel vector~\cite{loyka2001channel}.

Motivated by this idea, we start by evaluating the EM channel correlation at two separate points located in the receiver's region ${\bf x}, {\bf x'} \in V_{\rm R}\subset \mathbb{R}^3$. By assuming unit transmission power, the channel from the source to receiver is proportional to the received electric field observed at ${\bf x}$, i.e., ${\bf E}({\bf x})\in\mathbb{C}^3$, and the definition of ${\bf E}({\bf x'})$ is similar. Notice that our goal is to evaluate the correlation matrix ${\bf R}({\bf x}; {\bf x'}):=\mathbb{E}[{\bf E}({\bf x}){\bf E}\H({\bf x'})]$ with some reasonable assumption on the surrounding scattering conditions. Since an arbitrary EM field configuration can be accurately represented by an infinite superposition of planar waves with complex amplitude ${\bf E}_0({\bm \kappa})$~\cite{wan2023mutual}, i.e., 
\begin{equation}
    {\bf E}({\bf x}) = \frac{1}{(2\pi)^3}\int_{\mathbb{R}^3} {\bf E}_0({\bm \kappa})e^{\ri {\bm \kappa}\cdot {\bf x}} {\rm d}^3{\bm \kappa},
\end{equation} 
we know immediately that the correlation ${\bf R}({\bf x}, {\bf x'})$ depends intrinsically on the statistics of the planar wave component ${\bf E}_0({\bm \kappa})$. By assuming an independent complex Gaussian amplitude of this component and noticing ${\bm \kappa} \cdot {\bf E}_0=0$, we obtain the correlation matrix (up to a constant factor) as 
\begin{equation}
    {\bf R}({\bf x}; {\bf x'}) \propto \int_{\hat{\bm \kappa}\in S^2} ({\bf I} - \hat{\bm \kappa}\hat{\bm \kappa}\T)e^{\ri k_0 \hat{\bm \kappa}\cdot ({\bf x} - {\bf x'})} \nu(\hat{\bm \kappa}) \diff S,
    \label{eqn:initial_correlation_integral}
\end{equation}
where the integral is taken over the surface of the unit sphere $S^2$, $\hat{\bm \kappa}$ is the unit radius vector, $k_0=2\pi/\lambda_0$ is the wavenumber, and $\nu: S^2 \to \mathbb{R}_+$ is the angular power spectral density of the incident wave, measured in Watts per steradian per polarization. 
\begin{remark}
    The integrand ${\bf I}-\hat{\bm \kappa}\hat{\bm \kappa}\T$ automatically ensures the Fourier transformed version of Gauss's law for electric fields $\nabla\cdot {\bf E}=0$, i.e., $\ri {\bm \kappa}\cdot {\bf E}_0 =0$. 
\end{remark}
\red{By further assuming isotropic EM incidence, i.e., $\nu(\hat{\bm \kappa})\equiv \sigma^2/(8\pi)$, we directly compute the closed-form formula of ${\bf R}({\bf x}; {\bf x'})$ to be}
\begin{equation}
	\begin{aligned}
        {\bf R}({\bf x}; {\bf x'}) :=& {\mathbb E}\left[{\bf E}(\bf x){\bf E}\H (\bf x')  \right]\\
		  = &\frac{\sigma^2}{8}(f_0(k_0 r)+f_2(k_0 r)){\bf{I}}_3 \\
        & + \frac{\sigma^2}{8}(f_0(k_0 r)-3f_2(k_0 r)){\bf{\hat{r}}}{\bf{\hat{r}}\T },
	\end{aligned}
    \label{eqn:EM_correlation}
\end{equation} 	
where the radius vector is defined as ${\bf r} = {\bf x}-{\bf x'}$, $r = \sqrt{{\bf r}\T {\bf r}}$, $\hat{\bf r} = {\bf r}/r$, and the single-variable analytic function $f_n(\beta)$ is given by~\cite{hormander1973}
\begin{equation}
        f_n(\beta) = \int_{-1}^{1}x^n e^{\ri \beta x}\diff x, 
        \label{eqn:def-fn-function}
\end{equation}
which is the (inverse) Fourier transform of the truncated power function $x^n \mathbbm{1}_{[-1, 1]}(x)$. The proof of~\eqref{eqn:EM_correlation} is shown in {\bf Appendix~\ref{app:proof_correlation_integral}}. 

\begin{remark}
    The EMCF~\eqref{eqn:EM_correlation} is expressed in an explicit form under the isotropic incidence assumption $\nu\propto 1$, describing the correlation factor between a pair of antennas located in arbitrary positions ${\bf x}$, ${\bf x'}$ and arbitrary orientations. Given the unit polarization vector ${\bf p}$, ${\bf p'}\in S^2\subset\mathbb{R}^3$, the correlation of the measured channel responses of this pair of antennas is given by ${\bf p}\T {\bf R} {\bf p'}\in\mathbb{C}$. 
\end{remark}

\begin{remark}
    The EMCF~\eqref{eqn:EM_correlation} is invariant under spatial translations, since the correlation values only depend on the spatial displacement vector ${\bf r} = {\bf x} - {\bf x'}$. Thus, Eqn.~\eqref{eqn:EM_correlation} can be written in a more compact form ${\bf R}({\bf r})$. 
\end{remark}

Note that the foregoing coefficient $\sigma^2/8$ in~\eqref{eqn:EM_correlation} ensures the channel energy normalization condition $\tr({\bf R}({\bf x}; {\bf x})) = \sigma^2, \,\forall {\bf x}\in {\mathbb{R}}^3$. Interestingly, although the correlation model~\eqref{eqn:EM_correlation} is obtained under isotropic scattering conditions, it can be easily extended to the non-isotropic case by the following simple variable substitution:
\begin{equation}
    {\bf r} \to {\bf r} - \ri {\bm \mu}/k_0. 
    \label{eqn:analytic-continuation}
\end{equation}
This is due to the fact that replacing ${\bf r}$ by ${\bf r} - \ri {\bm \mu}/k_0$ is equivalent to introducing an exponential modulation factor to $\nu$, rendering $\nu = (\sigma^2/(8\pi))e^{\hat{\bm \kappa}\cdot {\bm \mu}}$ to the integrand in~\eqref{eqn:initial_correlation_integral}. This modulation factor $e^{\hat{\bm \kappa}\cdot {\bm \mu}}$ introduces non-isotropic properties to the incoming EM waves by assigning larger weights to the directions parallel to ${\bm \mu}$, as well as suppressing the directions anti-parallel to ${\bm \mu}$. Thus, we call ${\bm \mu} \in {\mathbb{R}}^3$ the {\it concentration parameter}. Moreover, the direction of ${\bm \mu}$ indicates the EM-concentrated direction, and the norm $\|{\bm \mu}\|$ is the concentration. Note that the larger the concentration is, the sparser the channel will be. 
\red{This ${\bm \mu}$-parameterized distribution on $S^2$ is also known as the von Mises-Fisher (vMF) distribution~\cite{mardia1976bayesian}, which has been widely applied to model the angles of departure (AoDs) and angles of arrival (AoAs) of wireless channels~\cite{jaeckel2014quadriga,CDL,mammasis2009spatial}.} 

The variable substitution in~\eqref{eqn:analytic-continuation} is well-known as the analytic continuation technique~\cite{ahlfors1979complex}, where the domain of a real-variable function is extended to complex values and maintains favourable analytic properties such as complex differentiability and the validity of closed-form expressions. This analytic continuation process is analogous to the extension from Fourier transforms to Laplace transforms, where Fourier harmonic integral kernel $e^{-\ri \omega t}, \omega\in\mathbb{R}$ is replaced by its analytic continuation $e^{-st},\,s\in \mathbb{C}$. By working in the complex domain, the concentration property can be seamlessly integrated into the existing correlation model~\eqref{eqn:initial_correlation_integral}. 
\begin{remark}
    It can be observed in~\eqref{eqn:analytic-continuation} that the real part of the complex displacement vector $\bf r$ means spatial separation of different receivers, while the imaginary part of $\bf r$, i.e., ${\bm \mu}$, corresponds to the directional concentration of incoming EM waves. By analytic continuation, these two different physical quantities can be treated equally as a single complex vector. 
\end{remark}

\red{Although the vMF assumption leads to a closed-form expression, one may argue that the assumption is too strong to be general. In fact, it has been established in~\cite{ng2022universal} that a {\it finite} mixture of $m$-dimensional von Mises-Fisher (vMF) distributions can $\mathcal{L}^\infty$-approximate arbitrary distributions with continuous pdfs on the $m$-dimensional sphere $S^m$, up to arbitrary distribution $\varepsilon$. Thus, for arbitrary electromagnetic incidence density $\nu(S)$, the density can be approximated by a finite number of EMCFs up to arbitrary specified precision $\varepsilon$. This finite approximation corresponds to the mixed-kernel EMCF, which will be introduced in the next section. As a result, the vMF assumption does not affect the generality of the proposed EMCF channel model. }

Despite the favorable mathematical properties brought by~\eqref{eqn:analytic-continuation}, the introduction of a non-constant modulation factor $\nu$ also violates the energy normalization condition $\tr({\bf R}) = \sigma^2$. Thus, we need to introduce an additional normalization factor
\begin{equation}
    C(\mu) = \int_{S^2}\tau(\hat{\bm \kappa}, {\bm \mu}){\rm d}S = \sinh(\mu)/\mu,
\end{equation}
where $\mu = \|{\bm \mu}\|\in\mathbb{R}_+$, and $\tau(\hat{\bm \kappa}, {\bm \mu}) = e^{\hat{\bm \kappa}\cdot {\bm \mu}}/(4\pi)$. By applying this normalization factor, the EMCF is written as 
\begin{equation}
    {\bf K}_{\rm EM}({\bf x}; {\bf x'}) = \frac{\sigma^2}{C(\|{\bm \mu}\|)} {\bf \Sigma}(k_0 {\bf z}),
    \label{eqn:Three-dimensional-SCF}
\end{equation}
where ${\bf K}_{\rm EM}$: $\mathbb{C}^3\to \mathbb{C}^{3\times 3}$ is called the electromagnetic correlation function (EMCF), i.e., the {\it EM kernel}. In the above expression of the EMCF, the complex displacement vector is defined as ${\bf z} = ({\bf x} - {\bf x'}) - \ri {\bm \mu}/k_0\in\mathbb{C}^3$, and the matrix-valued correlation function that appears above is written as 
\begin{equation}
    \begin{aligned}
    {\bf \Sigma}({\bf w}) &= \frac{1}{8}(f_0(|{\bf w}|)+f_2(|{\bf w}|)){\bf{I}}_3 \\
    & + \frac{1}{8}(f_0(|{\bf w}|)-3f_2(|{\bf w}|)){\bf{\hat{w}}}{\bf{\hat{w}}\T}.
    \end{aligned}
    \label{eqn:def_Sigma_of_w}
\end{equation}

\subsection{Extension to Fast Fading Channels}
Channel fading is the culprit of communication outages. Among the four types of channel fading, i.e., large-scale, small-scale, frequency-selective and time-selective fast fading\footnote{These non-ideal effects may occur at the same time. }, the last two are the most difficult to combat due to their rapid changing properties. 
The proposed EMCF is capable of explaining the spatial correlation of small-scale fading in an EM-compliant manner. To further describe the fast temporal channel variations, the spatial correlation function ${\bf R}({\bf x}; {\bf x'})$ should be extended into a spatio-temporal correlation function (STCF), which is expressed as ${\bf R}({\bf x}, t; {\bf x'}, t')$. Generally, this function has no analytical form. 

Fortunately, by assuming a von Mises density in the wavenumber domain, the analytical expression of the STCF with additional temporal variable $t$ can be obtained by the same analytic continuation technique. In fact, by introducing a velocity vector ${\bm v}\in {\mathbb{R}}^3$, we can incorporate time selectiveness, i.e., Doppler shift, into the EMCF. The EMCF is then augmented to support space-time variables $({\bf x}, t)\in\mathbb{R}^4$. In this case, the spatial-temporal correlation function is expressed as 
\begin{equation}
    \begin{aligned}
        {\bf K}_{\rm EM}({\bf x}, t; {\bf x'}, t') = \frac{\sigma^2}{C(\|{\bm \mu}\|)} {\bf \Sigma}(k_0 {\bf z}),
    \end{aligned}
    \label{eqn:Four-dimensional-STCF}
\end{equation}
where ${\bf z} = {\bf r} + {\bf v}(t-t')-\ri {\bm \mu}/k_0 \in \mathbb{C}^3$. 
\begin{remark}
    In Eqn.~\eqref{eqn:Four-dimensional-STCF}, it can be observed that the Doppler effect caused by the motion of the receiving antenna array is equivalent to an antenna displacement ${\bf v} (t-t')$. This conclusion is as expected, since the Doppler frequency shift is caused by the motion of the observer. In other words, the STCF~\eqref{eqn:Four-dimensional-STCF} automatically incorporates the Doppler effect. 
\end{remark}

\begin{figure*}[!t]
    \centering
    \includegraphics[width=1\linewidth]{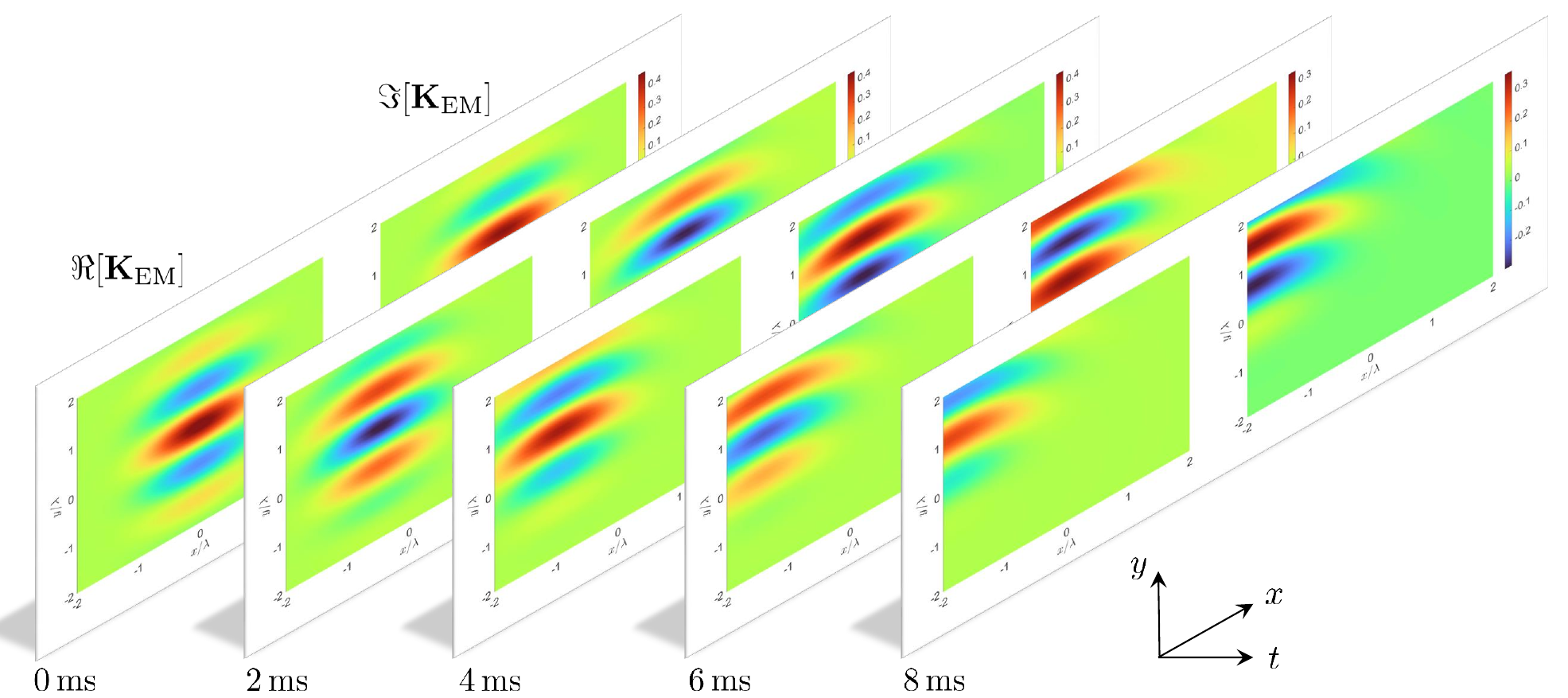}
    \caption{Slice chart of the space-time EMCF ${\bf K}_{\rm EM}({\bf r}, t)$, where ${\bf r} = (x, y, z)$. The real part and the imaginary part of the EMCF are shown in two horizontal groups. }
    \label{fig:EM-kernel-spacetime}
\end{figure*}

\subsection{Properties of the EMCF} \label{sec:3-3}
Since the EMCF ${\bf K}_{\rm EM}: \mathbb{R}^4\to \mathbb{C}^{3\times 3}$ is a four-dimensional complex-argument matrix-valued function, we study the mathematical properties from two aspects: (1) ${\bf K}_{\rm EM}$ as a function of the real argument, i.e., the spatial displacement vector ${\bf r} = {\bf x} - {\bf x'}$ and the time difference $\Delta t = t-t'$; (2) ${\bf K}_{\rm EM}$ as a function of the imaginary argument, i.e., the concentration parameter ${\bm \mu}$.

(1) {\it The EMCF v.s. real displacement $\bf r$.}  First, we assume $\Delta t=0$, i.e., we focus on the channel correlation between different antennas at the same instant. Since the real displacement vector ${\bf r}$ describes the vector pointing from the reference antenna to the observation antenna, the value of the EMCF is exactly the channel correlation coefficient between the reference antenna and the observation antenna, up to a constant scaling. These correlation coefficients are especially instructive in channel prediction tasks~\cite{liu2013mimo}, where only partial or previous channel coefficients are observable but complete channel matrices are required for subsequent signal processing. By evaluating the EMCF as a function of ${\bf r}$, a prior knowledge about the spatial correlation coefficients can be readily obtained. 

Second, we fix ${\bf r} = {\bf 0}$, and allow the time difference $\Delta t$ to vary. In this case, the EMCF describes the temporal correlation of the channel. It is worth noting that, the variation of the complex channel gain is attributed to the relative movement of the transmitter and the receiver. A classical temporal correlation model is the Jake's model (or Clarke's model)~\cite{jakes1974mobile,byers2004spatially}. In this model, the temporal channel correlation is expressed as 
\begin{equation}
    c_h(t, t'):=\mathbb{E}[h(t)h^*(t')] = \sigma_h^2 J_0(2\pi v(t-t')/\lambda),
\end{equation}
where $v\,{\rm [m/s]}$ is the velocity of the moving receiver, $\lambda\,[{\rm m}]$ is the operating wavelength, and $\sigma_h^2$ is the mean channel energy. The Jake's model is derived in a scalar wave 2-D settings with isotropic scattering, i.e., the incoming wave has no directivity~\cite{pizzo2020spatially}. \red{The Jake's model is exact in the scalar wave isotropic environment, however, it neglects the vector nature of EM fields. In contrast, the tri-polarized vector property and the non-isotropic property are well-captured by the proposed EMCF~\eqref{eqn:Four-dimensional-STCF}. Thus, the EMCF can be viewed as an electromagnetically compliant extension of the classical Jake's model. }

To summarize the above spatial and temporal discussions about the EMCF, we illustrate the EMCF as a function of the spacetime 4-vector $({\bf x}, t)$ in Fig.~\ref{fig:EM-kernel-spacetime}. In the simulation, the concentration vector ${\bm \mu} = [0, 10, 10]\T$, and the user's velocity is ${\bf v} = [20, 0, 20]\T\,{\rm m/s}$ with $y$-polarized antennas. It can be observed from Fig.~\ref{fig:EM-kernel-spacetime} that: (a) The spatial correlation peaks at the origin ${\bf r} = {\bf 0}, \Delta t=0$, and decays with oscillation as ${\bf r}\to\infty, \Delta t\to\infty$; (b) The velocity ${\bf v}$ causes a translation of the covariance in the spatial domain. \red{The first observation is consistent with existing channel correlation models in the literature~\cite{wang2022electromagnetic, pizzo2020spatially, byers2004spatially, abdi2000versatile}, but the second observation has not been revealed before. In fact, the temporal channel variation are usually attributed to Doppler frequency shift in the literature, but intrinsically it is caused by the movement of the receiving antenna. Thus, a unified EM model that combines the velocity ${\bf v}$ with the displacement ${\bf r}$ is self-complete, i.e., no additional Doppler models are needed.   }

(2) {\it The EMCF v.s. imaginary displacement $\bm \mu$.} Unlike the previously discussed ${\bf r}$, the concentration parameter ${\bm \mu}$ is often neglected in existing works. From the analytic continuation derivation of~\eqref{eqn:Three-dimensional-SCF} we can observe that, the introduction of ${\bm \mu}$ is equivalent to an additional spatial non-istropic factor $\nu \propto e^{\hat{\kappa}\cdot {\bm \mu}}$. This factor $\nu$ is mathematically equivalent to a 3D vMF angular distribution of the impinging waves~\cite{pizzo2022fourier, pizzo2020spatially}. However, the authors of~\cite{pizzo2022fourier} does not realize the analytic continuation technique, and thus they do not provide a closed-form expression of the correlation function. The existence of such a closed-form expression avoids the complicated evaluation of the coupling coefficients in the Fourier plane-wave domain, thus enabling low-complexity parameter fitting without optimizing for a large number of parameters as in ~\cite[Eqn.~(38)]{pizzo2022fourier}.

\begin{figure}[ht]
    \centering 
    \includegraphics[width=\widthRatio\linewidth]{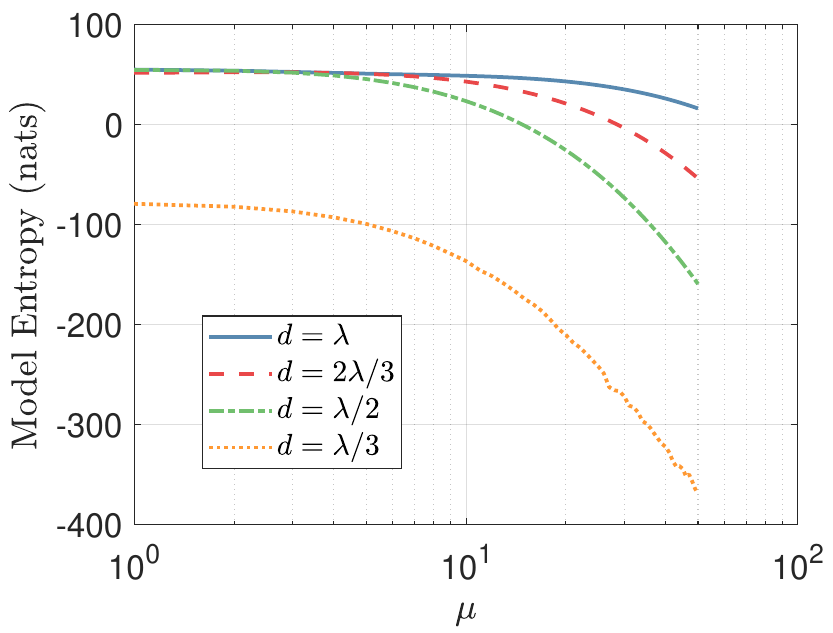}
    \caption{Kernel entropy as a function of the concentration parameter $\mu$ with different antenna spacing settings.  }
    \label{fig:kernel-entropy}
\end{figure}

Another important observation is that the imaginary displacement ${\bm \mu}$ controls the complexity of the EMCF. Due to spherical symmetry, the model complexity is independent of the direction $\hat{\bm \mu} := {\bm \mu}/\|{\bm \mu}\|$, but it depends on the scalar concentration $\mu := \|{\bm \mu}\|$. Fig.~\ref{fig:kernel-entropy} shows the relationship between model entropy and the concentration parameter $\mu$, with $\hat{\bm \mu} = [0, 1, 1]\T/\sqrt{2}$, $N_{\rm BS} = 12$, and various antenna spacings $d$. The model (differential) entropy $H(\mu)$ is defined as $\log\det(\pi e {\bf K})$~\cite{cover1999elements}, \red{where ${\bf K}\in \mathbb{S}_{+}^{N_{\rm BS}}$ is particularized from the EMCF ${\bf K}_{\rm EM}$ with spatial spacing $d$ and polarization aligned to the $y$-axis}. From Fig.~\ref{fig:kernel-entropy}, it can be concluded that the proposed EMCF becomes less complicated, i.e., contains less parameters, when $\mu\to\infty$. This is because highly concentrated impinging waves are closer to planar waves, while planar waves are the simplest field configurations that contain the least number of parameters. \red{In fact, planar waves can be naturally recovered from the EMCF by taking $\mu\to \infty$. To be specific, we can prove that 
\begin{equation}
    \lim_{\rho\to\infty} {\bf K}_{\rm EM}({\bf x}-\ri \rho\hat{\bm \mu}/k_0; {\bf x'}) = \frac{\sigma^2}{2}({\bf I} - \hat{\bm \mu}\hat{\bm \mu}\T)e^{\ri k_0 \hat{\bm \mu}\cdot ({\bf x-x'})}, 
\end{equation}
where $\hat{\bm \mu}\in S^2$ is fixed. Physically speaking, the right hand side of the limit explains the polarization and spatial phase difference of a strong line-of-sight (LoS) path composed of a single planar wave, i.e., the proposed EMCF explains strong LoS paths as a natural limit case. }

\subsection{Comparison with Other Spatially Correlated Channel Models}
\red{The proposed EMCF is intrinsically based on a spatially correlated Rayleigh fading channel model~\cite{bjornson2017massive}, while the novelty of this paper lies in that our spatial correlation function~\eqref{eqn:Three-dimensional-SCF} is both electromagnetically compliant and readily applicable to channel estimation. Most of the spatially correlated channel models assume that the channel vectors are circularly symmetric Gaussian distributed\footnote{Spatially correlated models are usually adopted to explain the small-scale fading effect of the wireless channel, which corresponds to the non-line-of-sight (NLoS) channel components. The LoS components are modeled in a location-determined way with deterministic parameters. } according to some predefined covariance matrix. This kind of channel models are traditionally applied to link-level simulations~\cite{mcnamara2002spatial,kammoun2015generalized,forenza2007simplified} and capacity evaluation tasks~\cite{chiani2003capacity,byers2004spatially}. Some recent works reveal that spatially correlated channel models are beneficial to MMSE channel estimators~\cite{bacci2024mmse,demir2022channel}, which indicates that accurate covariance modeling is a prerequisite for the construction of channel estimators.  
However, due to the recent academic interest of revisiting the EM physical foundations of wireless communications~\cite{migliore2018horse} via {\it EIT}, researchers have discovered that some of the existing spatial correlation models~\cite{forenza2007simplified,mcnamara2002spatial} do not satisfy the EM Helmholtz equation~\eqref{eqn:Helmholtz_equation}, implying that the model accuracy is not theoretically guaranteed. As a result, some  recent works are focused on constructing EM-compliant correlated channel models~\cite{pizzo2020spatially}. In this paper, apart from satisfying the EM constraints, we are further devoted to constructing {\it readily applicable} correlation functions that are beneficial to wireless-related statistical inference tasks. This requires the spatially correlated channel model to be simple with nice closed-form expressions. As a result, it is meaningful to propose the EMCF~\eqref{eqn:Four-dimensional-STCF} with an analytically compact expression in order to facilitate the subsequent model-based parameter fitting and channel inference. 
}

\section{Proposed EIT-MMSE Channel Estimation Algorithm} \label{sec4-Proposed-EIT-GPR-Channel-Estimation-Algorithm}
In this section, we will propose a novel Gaussian process regression (GPR)-inspired modified MMSE channel estimator based on the proposed EMCF. 

\subsection{Gaussian Process Regression (GPR)}
Gaussian process regression (GPR) methods can provide prediction from a GRF prior and the observed data~\cite{williams2006gaussian}. The GRF is specified by $f(x) \sim \mathcal{GRF}(m(x), k(x,x'))$, and the data $y_i$ is observed on a set of points $\{x_i\}_{i=1}^N\subset D$, satisfying $y_i = f(x_i) + \epsilon_i$. The observation noise $\epsilon_i \overset{\rm i.i.d}{\sim} \mathcal{CN}(0, \sigma_\epsilon^2)$. The objective of GPR is to generate prediction $f_*:=f(x_*)$ at any $x_*\in D$, given the observed data ${\bf y} = \{y_i\}_{i=1}^N$.  

Since the mean and covariance of the joint observation-prediction vector ${\bf z} = (y_1, y_2, \cdots, y_N, f_*)\T\in\mathbb{C}^{N+1}$ is given by 
\begin{equation}
    \begin{aligned}
        {\bf m} &=  (m(x_1), m(x_2), \cdots, m(x_N), m(x_*))\T, \\
        {\bf C} &= \begin{bmatrix}
            k(x_1, x_1) & k(x_1, x_2) & \cdots & k(x_1, x_*) \\
            k(x_2, x_1) & k(x_2, x_2) & \cdots & k(x_2, x_*) \\
            \vdots      & \vdots      & \ddots & \vdots      \\
            k(x_N, x_1) & k(x_N, x_2) & \cdots & k(x_N, x_*) \\
            k(x_*, x_1) & k(x_*, x_2) & \cdots & k(x_*, x_*) \\
        \end{bmatrix},
    \end{aligned}
    \label{eqn:GPR-joint-mean-and-variance}
\end{equation}
thus by applying the Gaussian posterior formula~\cite{wytock2013sparse}, we obtain
\begin{equation}
    \begin{aligned}
        {m}_{f_*|{\bf y}} &= m(x_*) + {\bf k}\H {\bf C}_{\bf y}^{-1}({\bf y} - {\bf m}_{\bf y}) \in \mathbb{C}, \\
        {c}_{f_*|{\bf y}} &= k(x_*, x_*) - {\bf k}\H {\bf C}_{\bf y}^{-1} {\bf k} \in \mathbb{R}_{+},  
    \end{aligned}
    \label{eqn:GPR-predictive-mean-and-variance}
\end{equation}
where
\begin{equation}
    \begin{aligned}
        {\bf k} &= {\bf C}(1:N, N+1) \in \mathbb{C}^{N\times 1}, \\
        {\bf C}_{\bf y} &= {\bf C}(1:N, 1:N) + \sigma_\epsilon^2 {\bf I}_N\in \mathbb{C}^{N\times N}. 
    \end{aligned}
\end{equation}
The predictive mean, i.e., the regression result, is given by $m_{f_*|{\bf y}}$ in~\eqref{eqn:GPR-predictive-mean-and-variance}, together with the predictive variance $c_{f_*|{\bf y}}$. Since the prior distribution is Gaussian, the Bayesian GPR estimator coincides exactly with the maximum {\it a posteriori} (MAP) estimator and the maximum likelihood (ML) estimator. Due to its Bayesian optimality, the GPR predictor has been widely applied to various high-precision inference tasks including curve fitting, pattern recognition and data interpolation. Furthermore, in addition to the Bayesian optimality, the wide application of GPR is also attributed its capability of tuning the EMCF (kernel) $k(x,x')$, which will be explained in detail in the next subsection. 

\subsection{Kernel Learning} \label{sec:4-2}
The GPR kernel $k(x;x')$ encodes the prior knowledge of the Gaussian random field $f({\bf x})$ in an implicit manner, i.e., the function $f({\bf x})$ is not written in an explicit formula. This implicit characteristics allow more field configurations to occur, and thus lead to enhanced model expressibility. To further enable more model flexibility, we usually assume a tunable kernel $k(x;x' | \theta)$ with $\theta \in \Theta\subset \mathbb{R}^n$ being a tunable hyperparameter. By tuning this hyperparameter, the model is better fit to the observed data, and thus the prediction quality can be further improved. The procedure of finding the optimal hyperparameter is termed as {\it kernel learning}. 

Before finding the hyperparameter $\theta$, we must specify a criterion that evaluates which parameter is ``better'', given the observed data. Usually, the ML criterion is applied, i.e., 
\begin{equation}
    \hat{\theta}^{\rm ML} = \mathop{\arg\max}_{\theta \in \Theta} \log p({\bf y} | \theta),
\end{equation}
where the probability density function $p(\cdot | \theta)$ is given by 
\begin{equation}
    p({\bf y} | \theta) = \frac{1}{\pi^N \det {\bf C}_{\bf y}} \exp\left(-{\bf y}\H {\bf C}_{\bf y}^{-1} {\bf y}\right).
\end{equation}
Note that ${\bf C}_{\bf y} = {\bf C}_{\bf y}(\theta)$ is a function of the hyperparameter. In order to obtain the ML estimator $\hat{\theta}^{\rm ML}$, we need to compute the derivative of the log likelihood function $\ell(\theta | {\bf y}) := \log p({\bf y}|\theta)$ with respect to its hyperparameter $\theta$, and this derivative is expressed as 
\begin{equation}
    \begin{aligned}
    \frac{\partial \ell(\theta | {\bf y})}{\partial \theta_i} & = \frac{\partial}{\partial \theta_i}\left(-{\bf y}\H {\bf C}_{\bf y}^{-1} {\bf y} - \log\det {\bf C}_{\bf y}\right)\\
    &= {\bm \alpha}\H \frac{\partial {\bf C}_{\bf y}}{\partial \theta_i} {\bm \alpha} - {\rm tr}\left({\bf C}_{\bf y}^{-1} \frac{\partial {\bf C}_{\bf y}}{\partial \theta_i}\right)\\
    &= {\rm tr}\left(({\bm \alpha}{\bm \alpha}\H - {\bf C}_{\bf y}^{-1})\frac{\partial {\bf C}_{\bf y}}{\partial \theta_i}\right)
    \end{aligned}
    \label{eqn:GPR-ML-derivative-general}
\end{equation}
where ${\bm \alpha} = {\bf C}_{\bf y}^{-1}{\bf y}$, and $i=1,2,\cdots, n$ for each component $\theta_i$ of the hyperparameter $\theta$. Note that the above formula is only valid for real hyperparameters $\theta\in\mathbb{R}^n$. For complex hyperparameters, the derivatives $\partial/\partial\theta_i$ is replaced by Wirtinger derivatives $({\partial}/{\partial \theta_{i, {\rm Re}}} - \ri {\partial}/{\partial \theta_{i, {\rm Im}}})/2$~\cite{remmert1991theory}, and the derivative formula~\eqref{eqn:GPR-ML-derivative-general} remains unchanged due to the analyticity of $\ell$ w.r.t. the elements of ${\bf C}_{\bf y}$. Any gradient-based optimizer can be applied to obtain an approximation of $\hat{\theta}^{\rm ML}$. Unfortunately, for most forms of the kernel $k(x;x'|\theta)$, the entry of the matrix $[{\bf C}_{\bf y}]_{jk} = k(x_j, x_k) + \sigma^2 \delta_{jk}$ is not a convex function of $\theta_i$. Thus, generally speaking, the global optimality of a gradient optimizer applied to $\ell(\theta|{\bf y})$ is not guaranteed.

\subsection{Proposed EIT-MMSE Channel Estimator}
\begin{figure*}
    \centering
    \includegraphics[width=0.95\linewidth]{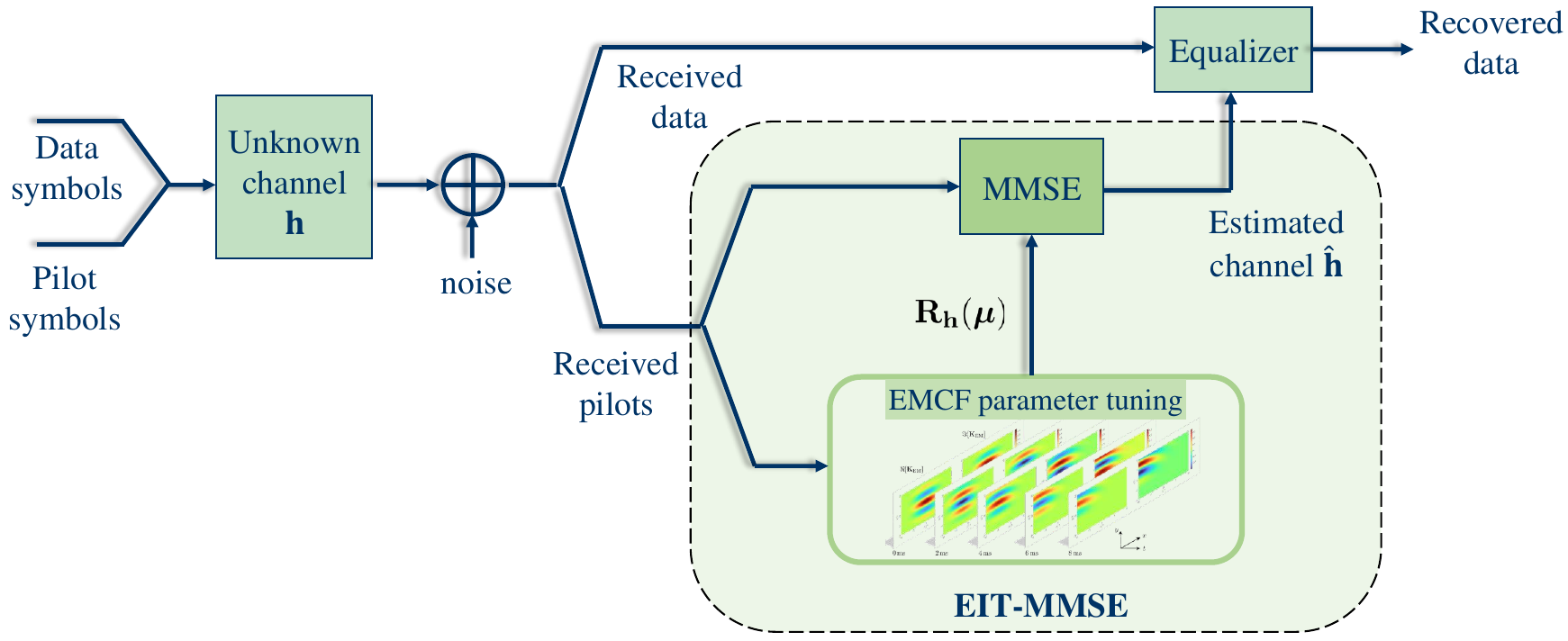}
    \caption{Schematic of the EIT-MMSE channel inference framework combined with EIT-Cov parameter tuning. The EMCF is first tuned by observed pilot vector ${\bf y}$. Then, the correlation matrix generated by EMCF is fed into the classical MMSE channel estimator to obtain the channel estimates. }
    \label{fig:EIT-CE-schematic}
\end{figure*}
We assume the BS is equipped with $N_{\rm BS}$ antennas placing at $\{{\bf d}_n\}\subset V_{\rm BS}\subset\mathbb{R}^3, n\in [N_{\rm BS}]$, where $V_{\rm Tx}$ is the spatial region that encloses the BS transmitter. The uplink channel is measured at time $t_p, \, p\in [P]$, with $P$ being the number of temporal snapshots. Following the GPR model, we can write the spacetime correlation tensor ${\bf R}$ between the $i$-th polarization at the $m$-th antenna and the $j$-th polarization at the $n$-th antenna as 
\begin{equation}
    {\bf R}_{mn,pq, ij} = {\bf p}_i\T[{\bf K}_{\rm EM}({\bf d}_m, t_p; {\bf d}_n, t_q)]{\bf p}_j,
\end{equation}
where ${\bf p}_i\in \mathbb{R}^3$ is the polarization direction of the antenna, 
$i, j\in [3]$, and ${\bf R}$ is a tensor with six indices. Although the correlation tensor ${\bf R}$ seems complicated, it contains various symmetric properties. One can verify that by simultaneously performing the following three index exchanges, the tensor component is conjugated:
\begin{equation}
    \begin{aligned}
        p &\leftrightarrow q, \, m\leftrightarrow n, \, i\leftrightarrow j. \\
    \end{aligned}
\end{equation}
Since the indices $(m, p, i)$ and $(n, q, j)$ always appear together, we introduce the compound index $\alpha = (m, p, i)$ and $\beta = (n, q, j)$ for notation clarity. With this notation, the displacement vectors ${\bf d}_m, {\bf d}_n$ are replaced by ${\bf d}_\alpha, {\bf d}_\beta$, and similar notation convention is applied to time instants $t$ and polarization vectors ${\bf p}$. 

The channel inference algorithm that takes in noisy channel observations and gives out estimated channel coefficient at arbitrary spatio-temporal coordinates is summarized in {\bf Algorithm~\ref{alg:Channel-inference}}. The key idea of this algorithm is to treat the unknown channel as a CSGRF, and then prediction can be performed by applying GPR. The algorithm framework is based on the computation of the posterior mean~\eqref{eqn:GPR-predictive-mean-and-variance}, and this is intrinsically equivalent with the MMSE estimator~\eqref{eqn:standard-MMSE-estimator} with channel correlation matrix ${\bf R}_{\bf h}$ replaced by the EMCF values. A schematic of the channel inference procedure is shown in Fig.~\ref{fig:EIT-CE-schematic}, together with other collaborating communication modules needed for data transmission. 

\begin{algorithm}[!t] 
	\caption{Proposed EIT-MMSE Channel Estimator} \label{alg:Channel-inference}
    \setstretch{1.2}
	\begin{algorithmic}[1]
		\REQUIRE
		Hyperparameters $\theta\in\Theta$; Received pilots $y_\alpha$, $\alpha\in A$; Noise variance $\sigma_\epsilon^2$; Predictive spacetime indices $\beta\in B$. 
		\ENSURE 
		Estimated channel coefficients $\hat{h}_\beta, \beta\in B$. \vspace{3pt}
		\STATE {\it \# Stage 1 (EMCF Evaluation):} 
        \STATE Let ${\bf K} \in \mathbb{C}^{|A|\times |A|}$, ${\bf y}\in\mathbb{C}^{|A|\times 1}$. 
		\FOR{$\alpha\in A$}
            \FOR{$\alpha' \in A$}
                \STATE Construct the EMCF: Compute ${\bf K}_{\alpha\alpha'} \leftarrow {\bf p}_\alpha\T{\bf K}_{\rm EM}({\bf x}_\alpha, t_\alpha; {\bf x}_{\alpha'}, t_{\alpha'}|\theta){\bf p}_{\alpha'} + \sigma_\epsilon^2 \delta_{\alpha\alpha'}$ by~\eqref{eqn:Four-dimensional-STCF}. 
            \ENDFOR
		\ENDFOR
        \STATE ${\bf a} \leftarrow {\bf K}^{-1}{\bf y}$. 
		\vspace{4pt}
		\STATE {\it \# Stage 2 (GPR Prediction):}
        \STATE Let ${\bf W}\in \mathbb{C}^{|B|\times |A|}$. 
		\FOR{$\beta\in B$}
            \FOR{$\alpha\in A$}
                \STATE ${\bf W}_{\beta\alpha} \leftarrow {\bf p}_\beta\T{\bf K}_{\rm EM}({\bf x}_\beta, t_\beta; {\bf x}_\alpha, t_\alpha|\theta){\bf p}_\alpha$ by~\eqref{eqn:Four-dimensional-STCF}. 
            \ENDFOR
        \ENDFOR
		\STATE Channel reconstruction: $\hat{\bf h} \leftarrow {\bf W}{\bf a}$. 
		\vspace{3pt}
		\RETURN Estimated channel $\hat{\bf h}$.  
	\end{algorithmic}
\end{algorithm}

\subsection{EIT Covariance Estimator with Kernel Learning}\label{sec:4-4}
\red{
In order to further improve the fixed-kernel EIT-MMSE channel inference routine in {\bf Algorithm~\ref{alg:Channel-inference}}, the problem of kernel learning remains to be solved. As is explained in Section~\ref{sec:4-2}, traditional GPR kernel learning requires evaluating the derivatives of the kernel value w.r.t. the hyperparameters. As for the EMCF, the kernel value is given by ${\bf p}_\alpha\T {\bf K}_{\rm EM} {\bf p}_\beta$, and the hyperparameters are the concentration parameter ${\bm \mu}$, the channel variance $\sigma^2$, and the Doppler velocity ${\bf v}$. } Generally speaking, the EMCF supports a spatio-temporal four-dimensional (4D) channel inference. However, in this paper, we only focus on the channel inference among spatial indices, i.e., MIMO channel estimation among multiple antennas. Thus, the hyperparameters are reduced to ${\bm \mu}$ and $\sigma^2$. Further spatio-temporal 4D channel estimation and inference are left for future works. 

The derivatives of the EMCF w.r.t. the reduced hyperparameters are summarized in the following {\bf Lemma~\ref{lemma:derivative-EM-kernel}}.
\begin{lemma}[Derivative of the EMCF] \label{lemma:derivative-EM-kernel}
    The Wirtinger derivative of ${\bf K}_{\rm EM}$ w.r.t. ${\bm \mu}(k)$ and $\sigma^2$ is respectively given by 
    \begin{equation}
        \begin{aligned}
            \frac{\partial {\bf K}_{\rm EM}}{\partial {\bm \mu}(k)} &= -\frac{\sigma^2}{C(\mu)} \left[\ri \frac{\partial {\bf \Sigma}({\bf w})}{\partial {\bf w}(k)} + \frac{C'(\mu){\bm \mu}(k)}{C(\mu)\mu} {\bf \Sigma}({\bf w})\right], \\
            \frac{\partial {\bf K}_{\rm EM}}{\partial (\sigma^2)} &= \frac{1}{C(\mu)}{\bf \Sigma}({\bf w}), \\
        \end{aligned}
    \end{equation}
    where ${\bf w} = k_0 {\bf z}$, ${\bf z} = ({\bf x}_\alpha - {\bf x}_\beta) - \ri {\bm \mu}/k_0$, $\mu = \|{\bm \mu}\|$, and
    \begin{equation}
        \begin{aligned}
            \frac{\partial {\bf \Sigma}({\bf w})}{\partial {\bf w}(k)}&= \frac{1}{8}\left[\ri (f_1 + f_3)\hat{\bf w}(k) {\bf I}_3 + \ri(f_1-3f_3)\hat{\bf w}(k) \hat{\bf w}\hat{\bf w}\T\right. \\
            & + \left. (f_0 -3f_2) (\partial_k \hat{\bf w}\cdot \hat{\bf w}\T + \hat{\bf w} \cdot\partial_k \hat{\bf w}\T) \right]. 
        \end{aligned}
        \label{eqn:derivative-Sigma-wrt-w}
    \end{equation}
    Here, the symbol $\hat{\bf w} = {\bf w}/|{\bf w}|$. 
\end{lemma}
\begin{IEEEproof}
    Details are shown in {\bf Appendix~\ref{app:proof-lemma-1}}. 
\end{IEEEproof}

\vspace{+2pt}

Notice that the goal of kernel learning is to fit the EMCF-defined GRF to the observed data ${\bf y}\in\mathbb{C}^{|A|\times 1}$. Before performing kernel learning, we first need to specify the marginal log likelihood function $\ell(\theta | {\bf y})$ as the objective function. To enhance the expressibility of the model, we choose the mixed EMCF to be the convex combination of multiple sub-correlation functions, i.e., 
\begin{equation}
    k_{\rm EM,Mixed}(x_\alpha;x_\beta|{\bm \theta}) = {\bf p}_\alpha\T\left(\sum_{s=1}^{S} w_s {\bf K}_{{\rm EM}}(x_\alpha;x_\beta|\theta_s)\right){\bf p}_\beta,
    \label{eqn:mixed-kernel-expression}
\end{equation}
where $w_s\geq 0$ are the non-negative weights satisfying $\sum_s w_s = 1$, and ${\bm \theta}$ represents the collection of all the hyperparameters that contain $S$ different hyperparameters $\theta_s\in\Theta$. Theoretically, a finite number of mixed vMF kernels can approximate the angular power spectrum of arbitrary incident EM fields. 

In the mixed kernel case, following~\eqref{eqn:GPR-ML-derivative-general}, the log likelihood function $\ell(\{{\bm \mu}_s\}_{s=1}^{S} | {\bf y})$ is expressed as 
\begin{equation}
    \begin{aligned}
        \ell(\{{\bm \mu}_s, w_s\}_{s=1}^S, \sigma^2 | {\bf y} ) &= \log p({\bf y} | \{{\bm \mu}_s, w_s\}_{s=1}^S)\\
        &=  -{\bf y}\H {\bf K}_{\bf y}^{-1}{\bf y} - \log\det{\bf K}_{\bf y} \\
        &~~+ {\rm const},
    \end{aligned}
    \label{eqn:mixed-kernel-objective-function}
\end{equation}
where ${\bf K}_{\bf y} = {\bf K}_{\bf h} + \sigma_\epsilon^2 {\bf I}_{N_A} = \sum_s w_s{\bf K}_s + \sigma_\epsilon^2 {\bf I}_{N_A}$, $N_A = |A|$, ${\bf h}\in\mathbb{C}^{|A|\times 1}$ represents the noiseless true values of the channel, and 
\begin{equation}
    ({\bf K}_{\bf h})_{\alpha\beta} = k_{\rm EM, Mixed}(x_\alpha; x_{\beta}|{\bm \theta}). 
\end{equation}

Notice that the objective function~\eqref{eqn:mixed-kernel-objective-function} is a function of ${\bf K}_{\bf y}$, which is further a matrix-valued function of hyperparameters ${\bm \theta} = (\{{\bm \mu}_s, w_s\}_{s=1}^{S}, \sigma^2)\subset \Theta$. Since the latter derivative $\partial {\bf K_y}/\partial {\bm \mu}_s(k)$ has been solved in {\bf Lemma~\ref{lemma:derivative-EM-kernel}}, we only need to compute the Wirtinger derivative of $\ell$ w.r.t. ${\bf K}_{\bf h}$. This is given by the following formula
\begin{equation}
    \frac{\partial \ell}{\partial {\bf K}_{\bf h}} = ({\bf a}{\bf a}\H - ({\bf K}_y^{-1}))^*, 
    \label{eqn:derivative-ell-wrt-Kh}
\end{equation}
where ${\bf a} = {\bf K}_{\bf y}^{-1}{\bf y}$. Combining~\eqref{eqn:derivative-ell-wrt-Kh} with the conclusion of {\bf Lemma~\ref{lemma:derivative-EM-kernel}}, we arrive at the real-variable derivative as 
\begin{equation}
    \frac{\partial \ell}{\partial {\bm \mu}_s(k)} = 2 w_s \Re\left[{\rm tr}\left( \frac{\partial {\bf K}_{\bf h}}{\partial {\bm \mu}_s(k)} ({\bf a}{\bf a}\H - {\bf K}_{\bf y}^{-1}) \right)\right], 
    \label{eqn:der-ell-wrt-mu_s(k)}
\end{equation}
and
\begin{equation}
    \frac{\partial \ell}{\partial w_s} = 2\Re\left[{\rm tr}\left( {\bf K}_s({\theta}_s) ({\bf aa}\H - {\bf K}_{\bf y}^{-1}) \right)\right], 
    \label{eqn:der-ell-wrt-weights}
\end{equation}
which can be used to solve the following kernel learning problem with the ML criterion
\begin{equation}
    \hat{\bm \theta}^{\rm ML} = \mathop{\arg\max}_{\bm\theta} \ell({\bf y}|{\bm \theta}). 
\end{equation}
Given the derivatives~\eqref{eqn:der-ell-wrt-mu_s(k)}, the numerical optimizer can be chosen as gradient ascent, conjugate gradient ascent, or Armijo-Goldstein backtracking search~\cite{dennis1996numerical} . Note that the two terms appearing in~\eqref{eqn:mixed-kernel-objective-function} represent the data fitness and model complexity, respectively. Thus, minimization of the objective $\ell$ will automatically balance between the data fitness and model complexity.

\begin{algorithm}[!t] 
	\caption{EIT-Cov Algorithm for EMCF Parameter Tuning} \label{alg:EIT-Kernel-learning}
    \setstretch{1.2}
	\begin{algorithmic}[1]
		\REQUIRE
		Received pilots $y_\alpha$, $\alpha\in A$; Noise variance $\sigma_\epsilon^2$; Number of mixed kernels $S$; Maximum iteration number $N_{\rm iter}$. 
		\ENSURE 
		Estimated hyperparameters $\{\hat{\bm \mu}_s, \hat{w}_s\}_{s=1}^{S}, \hat{\sigma^2}$. 
        \vspace{3pt}
        \STATE $\hat{\sigma^2} \leftarrow 2\sum_{\alpha\in A}|y_\alpha|^2/(|A|\cdot (1+\sigma_\epsilon^2))$. 
        \STATE Let ${\bf K}_{\bf y} \in \mathbb{C}^{|A|\times |A|}$, and ${\bf y}\in\mathbb{C}^{|A|\times 1}$ containing input data from $y_\alpha, \alpha\in A$. 
        \STATE Set $t\leftarrow 0$, and initialize the hyperparameters $\{\hat{\bm \mu}_s^{(0)}, \hat{w}_s^{(0)}\}_{s=1}^S$. 
        \STATE Initialize the learning rates of Armijo-Goldstein's optimizer. 
        
        \FOR{$t=1,2,\cdots,N_{\rm iter}$}
            \STATE Construct the EMCF ${\bf K}_{\bf y}$ from hyperparameters $\{\hat{\bm \mu}_s^{(t-1)}, \hat{w}_s^{(t-1)}\}_{s=1}^S$ by~\eqref{eqn:Four-dimensional-STCF} and~\eqref{eqn:mixed-kernel-expression}. 
            \STATE ${\bf a} \leftarrow {\bf K}_{\rm EM}^{-1}{\bf y}$. 
            \FOR{$s=1,2,\cdots, S$}
                \STATE Compute $\frac{\partial \ell}{\partial {\bm \mu}_s(k)}, \, k=1, 2, 3$ from~\eqref{eqn:der-ell-wrt-mu_s(k)}. 
                \STATE Update ${\bm \mu}_s^{(t)}$ with $\frac{\partial \ell}{\partial {\bm \mu}_s(k)}$ by Armijo-Goldstein's optimizer. 
            \ENDFOR
            \FOR{$s=1,2,\cdots, S$}
                \STATE Compute $\frac{\partial \ell}{\partial w_s}$ from~\eqref{eqn:der-ell-wrt-weights}. 
                \STATE Update ${w}_s^{(t)}$ with $\frac{\partial \ell}{\partial w_s}$ by Armijo-Goldstein's optimizer. 
            \ENDFOR
        \ENDFOR
		\vspace{3pt}
		\RETURN Estimated hyperparameters $\{\hat{\bm \mu}_s^{(N_{\rm iter})}, \hat{w}_s^{(N_{\rm iter})}\}_{s=1}^S$, and $\hat{\sigma^2}$.  
	\end{algorithmic}
\end{algorithm}

\begin{table}[t] 
    \centering
    \begin{threeparttable}
        \caption{Computational Complexity of Different Uplink Channel Estimators} \label{tab:channel_estimator_complexity}
        \setstretch{1.3}
        \begin{tabular}{|l|c|} 
            \hline
            Algorithm                               & Complexity        
                                    \\
            \hline
            OMP~\cite{do2008sparsity}               & $\mathcal{O}(N_{A}^2 N_{\rm iter})$           \\
            \hline
            AMP~\cite{donoho2010message}            & $\mathcal{O}(N_{A}^2 N_{\rm iter})$           \\
            \hline
            FBS~\cite{haghighatshoar2018low}        & $\mathcal{O}(GN_{A}N_sN_{\rm iter})$      \\
            \hline
            Proposed EIT-MMSE                       & $\mathcal{O}(N_{A}^2)$                        \\
            \hline
            Proposed EIT-Cov                        & $\mathcal{O}(N_{A}^3 S N_{\rm iter})$         \\
            \hline
        \end{tabular}
    \end{threeparttable}
\end{table}

\red{The EMCF parameter tuning procedure produces covariance channel estimates from noisy channel observations. Thus, it is also called the EIT covariance estimator (EIT-Cov). The EIT-Cov algorithm is summarized in~{\bf Algorithm~\ref{alg:EIT-Kernel-learning}}, and its performance will be numerically evaluated in the next section.} In addition, here we present the complexity of the proposed EIT-MMSE channel estimator and the EIT-Cov algorithm in~{\bf Table~\ref{tab:channel_estimator_complexity}}, compared with classical compressed sensing (CS)-based channel estimators and the FBS covariance estimator. For the FBS estimator, the number of atoms in the sparsifying dictionary is denoted by $G$. Note that although EIT-Cov requires a cubic complexity due to matrix inversion, the parameter tuning is not always required for each frame of the channel estimate, thanks to the slow-varying property of the channel statistics. In this way, the complexity of the EIT-Cov statistical learning can be reduced in the time-avaraged sense. 

\subsection{Possible Improvement of the EIT-Cov Method}
\red{
Although being physically explainable and achieving better performance under some practical channel datasets, the proposed EIT-Cov estimator does not directly integrate wideband information. This difficulty is caused by the monochromatic assumption (or time-harmonic assumption) adopted by almost all of the spatial correlation works~\cite{kammoun2015generalized,forenza2007simplified,chiani2003capacity,mcnamara2002spatial,pizzo2022spatial,pizzo2020spatially}, where these models are intrinsically applicable to only narrow-band systems. To overcome this drawback, we propose an improved wideband version of the EIT-Cov estimator to exploit the similarity of the power angular spectrum (PAS)~\cite{kammoun2015generalized} across different subcarriers~\cite{ali2019spatial}. Consider an OFDM BS with $M$ subcarriers, where we assume that the incoming EM waves observed at all of the $M$ subcarriers share the same mixed-kernel PAS 
\begin{equation}
    \nu(\hat{\bm \kappa}) = \sum_{s=1}^S w_s \frac{\exp({\bm \mu}_s\cdot\hat{\bm \kappa})}{4\pi C(\|{\bm \mu}_s\|)}. 
\end{equation}
Thus, the joint hyperparameter learning problem upon observing the wideband pilot matrix ${\bf Y} := [{\bf y}_1, {\bf y}_2, \cdots, {\bf y}_M]\in\mathbb{C}^{N_{\rm BS}\times M}$ at all of the $M$ subcarriers is formulated as 
\begin{equation}
    {\bm \theta}^\star = \mathop{\arg\max}_{{\bm \theta}\in\mathsf{F}} \sum_{m=1}^{M} \left(-{\bf y}_m\H {\bf K}_{{\bf y}_m}^{-1}({\bm \theta}) {\bf y}_m -\log\det {\bf K}_{{\bf y}_m}({\bm \theta}) \right), \label{eqn:wideband_GPR_fitting}
\end{equation} 
where ${\bm \theta}^\star$ denotes the optimal solution of all the learnable concentration and kernel weight parameters $\{({\bm \mu}_s^\star, w_s^\star)\}_{s=1}^S$, ${\bf K}_{{\bf y}_m}$ denotes the EMCF particularized to the $m$-th subcarrier frequency, and the feasible set $\mathsf{F}$ denotes the additional regularization condition on the hyperparameters (e.g., it is reasonable to assume that $\|{\bm \mu}_s\|\leq 30, \,\forall s$ to ensure numerical stability in computing the EMCF with analytic continuation). The complexity of solving the wideband optimization problem~\eqref{eqn:wideband_GPR_fitting} is equal to solving $M$ narrowband problems, thus the total covariance estimation complexity is $\mathcal{O}(MN_{\rm A}^3 S N_{\rm iter})$, where $N_{\rm iter}$ denotes the number of iterations for gradient-based optimizer. In summary, although we will test the baselines and the proposed algorithms within a narrowband framework, the proposed method can be readily extended to the wideband case, which requires further investigation and is left for future works.  }

\section{Simulation Results} \label{sec5-Simulation-Results}
\red{In this section, we will present detailed simulation results, demonstrating the physical compliance and the online channel statistics learning capability of the EIT-Cov algorithm. To further verify the practical benefits of the EMCF statistical channel model, we show the improved performance of the EIT-MMSE channel estimation algorithm. }

\subsection{Physical Compliance of EIT-Cov}
\red{In this subsection, we verify the physical compliance of the EIT-Cov algorithm by fitting the EMCF to geometric channel observations ${\bf y}$.} Both the single-kernel and the bi-kernel case are studied, i.e., $S\in\{1, 2\}$. The carrier frequency is set to $f_c = 3.5\,{\rm GHz}$, and the number of BS antennas is set to $N_{\rm BS}=32$. 
The uplink channel is generated by (a) the geometric model with Friis transmission formula; (b) the SV model~\cite{meijerink2014physical}; (c) the TR 38.901 model~\cite{CDL} with CDL-A delay profile. The geometric model produces channel vector ${\bf h}$ by directly calculating the channel coefficient as 
\begin{equation}
    {\bf h}(n) = \frac{\lambda}{4\pi d_n^r} e^{\ri k_0 d_n},\, n\in [N_{\rm BS}],
\end{equation}
where $d_i$ is the geometric distance (in meters) from the $i$-th BS antenna to the user, and $r=1$ is the path loss exponent. Note that the positive phase shift, i.e., $+\ri k_0d_n$, is in agreement with the physicists' convention~\cite{bauck2018note} of Fourier transforms.

\begin{figure}[t]
    \centering
    \includegraphics[width=\widthRatio\linewidth]{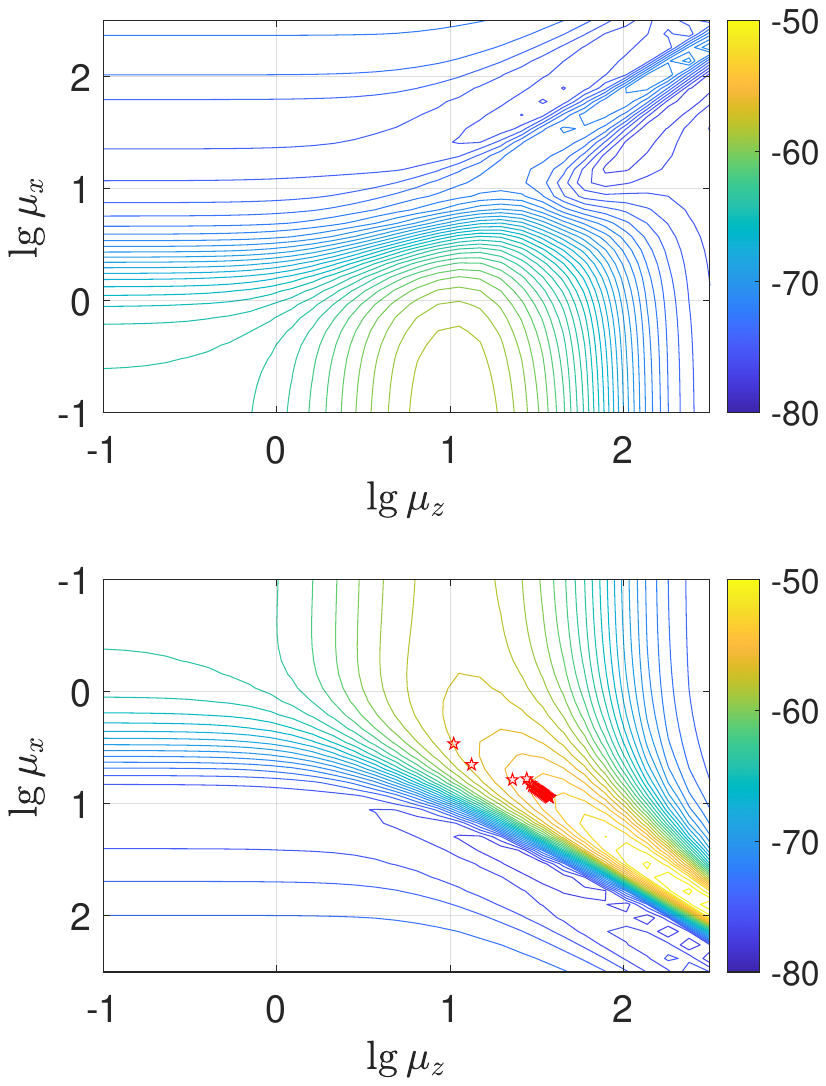}
    \caption{GPR objective function $\ln p({\bf y}|{\bm \mu}_{x,z})$ and trajectory of the Armijo-Goldstein backtracking optimizer (in red pentagrams). }
    \label{fig:GPR-objective-function}
\end{figure}

{\bf Simulation setup.} The EIT-Cov is studied by applying the geometric model (a) and tuning the kernel to fit the channel observations ${\bf y}$. Fig.~\ref{fig:GPR-objective-function} shows the contour plot of the single-kernel GPR objective function $\ell({\bm \mu}_x, {\bm \mu}_z|{\bf y})$. The variable ${\bm \mu}_y$ is fixed to zero, since the user is assumed to be located in the $xOz$ plane. During simulation, $\varphi_{\rm UE} = -15\deg$, $R_{\rm UE}=10\,{\rm m}$, and the receive SNR is set to be $\gamma=0\,{\rm dB}$. The user is intentionally placed in the near-field region of the BS array, in order to make the channel vector ${\bf h}$ non-DFT-like, i.e., the channel vector deviates from the far-field DFT channel. Although this non-DFT property prohibits traditional virtual channel-based angular analysis of the channel, this difficulty can be overcome by the proposed GPR channel fitter, since the EIT-Cov algorithm intrinsically supports multiple incoming waves. 

{\bf Parameter tuning performance}. The objective function $\ell({\bm \mu}_x, {\bm \mu}_z|{\bf y})$ is drawn in two parts in Fig.~\ref{fig:GPR-objective-function}. The upper half represents the objective function in the region $\{{\bm \mu}_x>0\}$, and the lower half represents the objective function in the region $\{{\bm \mu}_x<0\}$. For the sake of a broad functional vision, the variables are presented in log scale. The objective function attains better fitness at the lower half of the figure, and the region of better fitness extends in a ridge-shaped manner towards the southeast direction. In fact, this direction contains the information of the user's direction relative to the BS. Recall that the physical meaning of ${\bm \mu}$ is the concentration parameter. Thus, in a nearly line-of-sight propagation environment, the user's direction (azimuth) angle $\varphi_{\rm UE}$ satisfies
\begin{equation}
    \tan \varphi_{\rm UE} = \frac{{\bm \mu}_x}{{\bm \mu}_z}. 
\end{equation}
By applying the log transform, we obtain $\lg |{\bm \mu}_x| = \lg|\tan\varphi_{\rm UE}|+\lg|{{\bm \mu}_z}|$, meaning that the intercept of the ridge corresponds to the angular-related quantity $\lg\tan\varphi_{\rm UE}$. A coarse estimation of the interception produces $\lg|\tan\varphi_{\rm UE}|\approx 0.6$, which yields $\hat\varphi_{\rm UE}\approx -14\deg$. The accuracy of this estimation demonstrates the physical compliance of the proposed EMCF and the GPR-based EIT-Cov algorithm.  

\subsection{EIT-Cov Statistical Learning}
In this subsection, we numerically study the statistical learning performance of the proposed EIT-Cov covariance estimator by comparing it to the traditional sample covariance method as well as the FBS covariance estimator proposed in~\cite{haghighatshoar2018low}. 

{\bf Simulation setup.} Let $\hat{\bf R}_{\bf h}^{\rm ML}$ be the sample covariance matrix of the channel ${\bf h}$ computed from a noisy dataset $\{{\bf y}_i\}_{i=1}^{N_{s}}$ of $N_s$ samples, defined as $\hat{\bf R}_{\bf h}^{\rm ML} = (1/N_s)\sum_{i=1}^{N_s} {\bf h}_i {\bf h}_i\H - \sigma_n^2 {\bf I}$. The regularized sample covariance (SampleCovReg) $\hat{\bf R}_{\bf h}^{\rm reg}$ is constructed by replacing the negative eigenvalues of $\hat{\bf R}_{\bf h}^{\rm ML}$ with zeros, forcing $\hat{\bf R}_{\bf h}^{\rm reg}$ to be non-negative definite. 
\red{Let ${\hat{\bf R}_{\bf h}^{\rm FBS}}$ be the covariance estimate returned by the FBS algorithm~\cite[Alg.~2]{haghighatshoar2018low} when applied to the same noisy channel dataset. }
Let $\hat{\bf R}_{\bf h}^{\rm EIT}$ be the covariance estimated by fitting the EMCF to the same noisy dataset with {\bf Algorithm~\ref{alg:EIT-Kernel-learning}}. 
The dataset is constructed by calling the 3GPP channel model~\cite{CDL} in MATLAB communication toolbox. The true channel covariance ${\bf R}_{\bf h}^{\rm true}$ is obtained by first generating a noiseless channel dataset of size $N_s = 20,000$ and then computing the sample covariance. The average covariance estimation NMSE error is defined as $\mathbb{E}[\|\hat{\bf R}_{\bf h} - {\bf R}_{\bf h}^{\rm true}\|_{\rm F}^2/\|{\bf R}_{\bf h}^{\rm true}\|_{\rm F}^2]$, in which the expectation is approximated by Monte Carlo trials.

\begin{figure}[t]
    \centering
    \includegraphics[width=\widthRatio\linewidth]{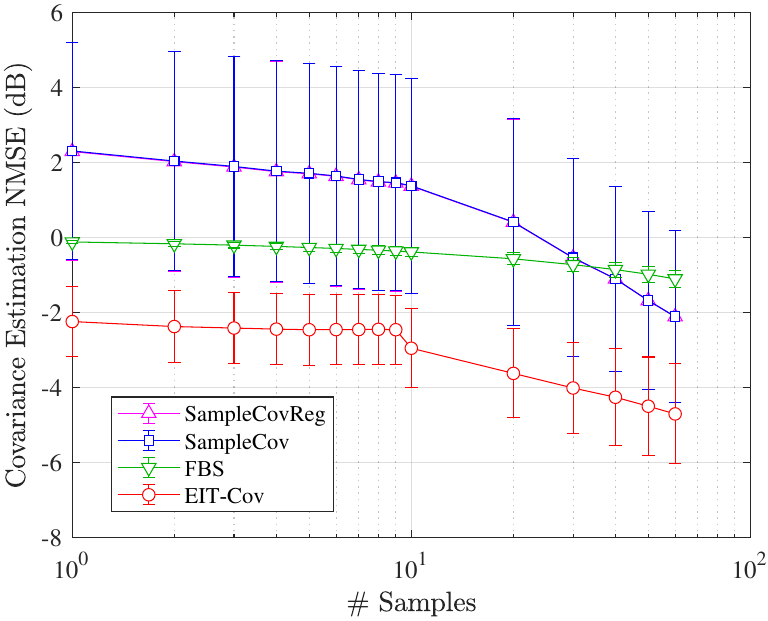}
    \caption{\red{Average covariance estimation performance of different channel statistics estimators. The SNR is set to $\gamma = 10\,{\rm dB}$. Error bars are computed from 1000 Monte Carlo trials. }}
    \label{fig:CovEst_Error_wrt_nSamples}
\end{figure}

\begin{figure}[t]
    \centering 
    \includegraphics[width=\widthRatio\linewidth]{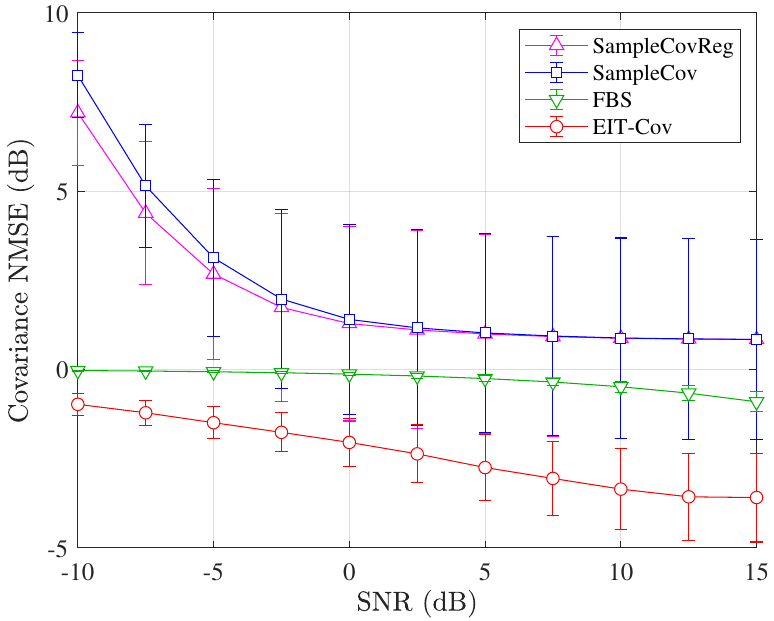}
    \caption{\red{Median covariance estimation performance of different channel statistics estimators. The dataset size is $N_s = 15$. Error bars are computed from 1000 Monte Carlo trials. }}
    \label{fig:CovEst_Error_wrt_SNR}
\end{figure}

\red{{\bf Covariance estimation performance.} Fig.~\ref{fig:CovEst_Error_wrt_nSamples} shows the covariance estimation error of $\hat{\bf R}_{\bf h}^{\rm ML}$, $\hat{\bf R}_{\bf h}^{\rm reg}$, $\hat{\bf R}_{\bf h}^{\rm FBS}$, and $\hat{\bf R}_{\bf h}^{\rm EIT}$ as a function of the number of samples $N_s$. When operating in the data-hungry regime $N_s \leq N_{\rm BS}=32$, the FBS covariance estimator achieves better NMSE performance than the standard sample covariance estimator (SampleCov), and the proposed EIT-Cov estimator outperforms FBS. When operating in the data-abundant regime $N_s > N_{\rm BS}$, the sample covariance method is generally better than FBS, however, the proposed EIT-Cov estimator still achieves the best NMSE performance compared with the FBS and sample covariance baselines.
Fig.~\ref{fig:CovEst_Error_wrt_SNR} shows the NMSE performance of covariance estimators as a function of SNR $\gamma$, with $N_s$ fixed to $15$. By exploiting the sparsity prior knowledge of FBS, the FBS covariance estimator generally achieves better performance than the sample covariance method. The proposed EIT-Cov estimator achieves the best NMSE performance across a wide range of SNR, demonstrating that the EIT-based prior knowledge is more beneficial than the sparsity-based prior knowledge when carrying out channel covariance estimation tasks on practical channel datasets. }

\subsection{GPR-based EIT-MMSE Channel Estimation}
\red{
In the previous Section~\ref{sec:4-4}, we have proposed a channel inference algorithm that exploits the prior information contained in the EMCF. Although {\bf Algorithm~\ref{alg:Channel-inference}} and {\bf Algorithm~\ref{alg:EIT-Kernel-learning}} are general, we apply them to the channel estimation task where the temporal variable $t$ is suppressed. Since for channel estimation problems the evidence set $A$ coincides with the prediction set $B$, we have $|A|=|B|=N_{\rm BS}$. Further extensions to channel prediction and space-time channel interpolation tasks are left for future works. }

{\bf Simulation setup.} \red{In the following channel estimation simulation, to ensure a more realistic channel property, we adopt (b) the standard SV multi-path synthetic channel model and (c) the standard 3GPP TR 38.901 CDL channel~\cite{CDL} for algorithm performance evaluation. Note that the channel generation mechanism does not favor the proposed estimators.} For the multi-path SV model, the number of NLoS paths is set to $L=6$, and the Rician factor is $K=10\,{\rm dB}$. The SV channel coefficients are normalized to ensure that $\mathbb{E}[\|{\bf h}\|^2] = N_{\rm BS}$. The SV LoS path is generated according to the geometric AoA, i.e., the physical angular direction $\varphi_{\rm UE}$ of the user, and the NLoS path AoAs are uniformly generated within range $[-\pi/2, \pi/2]$. For the 3GPP CDL model, we adopt the standard CDL-A delay profile.

{\bf Baseline algorithms. } The LS estimator is given by $\hat{\bf h}^{\rm LS} = {\bf y}$. The isotropic linear-MMSE (LMMSE-Iso) estimator is constructed by setting the MMSE correlation matrix ${\bf R} = {\bf K}_{\rm iso}$~\cite[Eqn.(8)]{demir2022channel} as 
\begin{equation}
    [{\bf K}_{\rm iso}]_{\alpha\beta} = \sigma^2 {\rm sinc}\left(\frac{2}{\lambda} \|{\bf x}_\alpha - {\bf x}_\beta \|\right). 
\end{equation}
The sample covariance methods first compute a covariance estimate from $N_s$ history channel samples, and then apply the standard LMMSE to obtain the channel estimate. Since the wireless multi-path channel exhibits angular sparsity, CS algorithms can be applied to exploit this sparse property, and thus achieve high-quality channel estimation. \red{In the simulation, we choose three methods as baselines, including the approximate message passing (AMP) algorithm, the orthogonal matching pursuit (OMP) algorithm, and the FBS estimator~\cite{haghighatshoar2018low}. The AMP algorithm is implemented according to~\cite{donoho2010message} with shrinkage parameter $\lambda=1.2$, and the OMP algorithm~\cite{do2008sparsity} is implemented with the number of path $L+1=7$, matching the SV channel generation procedure. For the FBS estimator, a number of $N_s$ covariance training samples are fed into the covariance estimator before performing standard MMSE channel estimation with the estimated channel covariance matrix.} All the channel estimators are evaluated by the normalized mean square error (NMSE) performance, which is defined as 
\begin{equation}
    {\rm NMSE} = \mathbb{E}\left[ \frac{\|\hat{\bf h} - {\bf h}\|^2}{\|{\bf h}\|^2}\right], 
\end{equation}
where the expectation $\mathbb{E}[\cdot]$ is calculated by Monte-Carlo simulations. 

\begin{figure}[t]
    \centering
    \includegraphics[width=\widthRatio\linewidth]{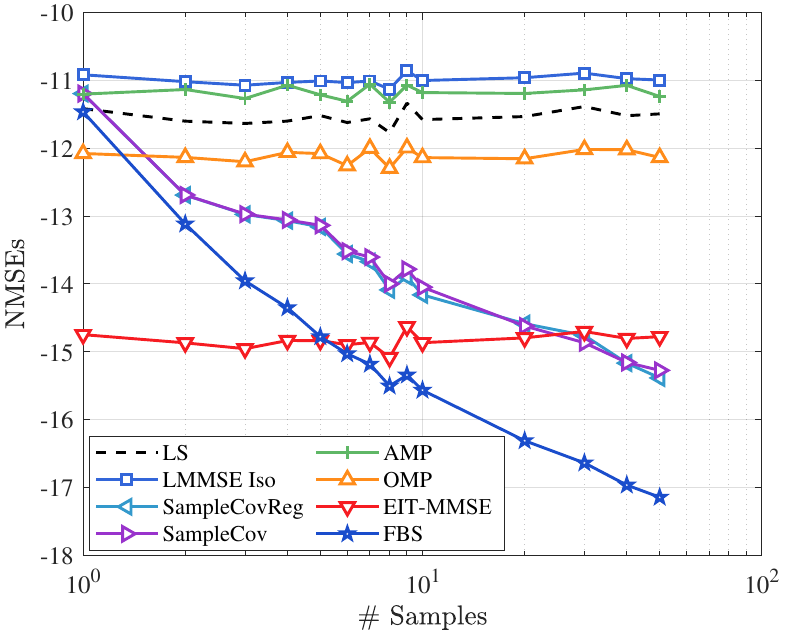}
    \caption{\red{NMSE performance of different channel estimators as a function of the number of training samples $N_s$, with fixed SNR $\gamma=10\,{\rm dB}$. The number of Monte Carlo trials is 1000. }}
    \label{fig:CE_NMSE_wrt_nSamples}
\end{figure}


{\bf Estimator performance w.r.t. $N_s$. } \red{We compare the NMSE performance of the traditional covariance estimation -- channel estimation procedure to that of the proposed EIT-MMSE channel estimator with EIT-Cov being a subroutine. Simulation results are shown in Fig.~\ref{fig:CE_NMSE_wrt_nSamples} with SNR $\gamma = 10\,{\rm dB}$. It is observed that the proposed EIT-Cov method outperform the sample covariance method when $N_s$ is less than 20. Different from the simulation of $\hat{\bf R}_{\bf h}^{\rm EIT}$, here the EIT-Cov channel estimation method operates in the {\it single-shot learning} regime, i.e., the EIT-Cov-based channel estimator $\hat{\bf h}^{\rm EIT}({\bf y})$ relies on the instant noisy observation ${\bf y}$ but not on the noisy historical dataset $\{{\bf y}_i\}_{i=1}^{N_{s}}$. The motivation of this assumption is to ensure fair comparison among the single kernel EIT-Cov method and the other single-shot channel estimators. It can be seen from Fig.~\ref{fig:CE_NMSE_wrt_nSamples} that the EIT-MMSE channel estimator outperforms all the baselines when $1\leq N_s \leq 5$, demonstrating that the EIT prior knowledge is worth about 5 noisy channel samples. When the number of available covariance training samples surpasses $N_{\rm BS}$, the sample covariance method becomes to outperform EIT-MMSE, which is because of the reduced singularity of sample covariance matrix in the data-abundant regime. In summary, the EM prior information is successfully embedded into the proposed EMCF and utilized by the MMSE channel estimator. }



\begin{figure}[t]
    \centering
    \includegraphics[width=\widthRatio\linewidth]{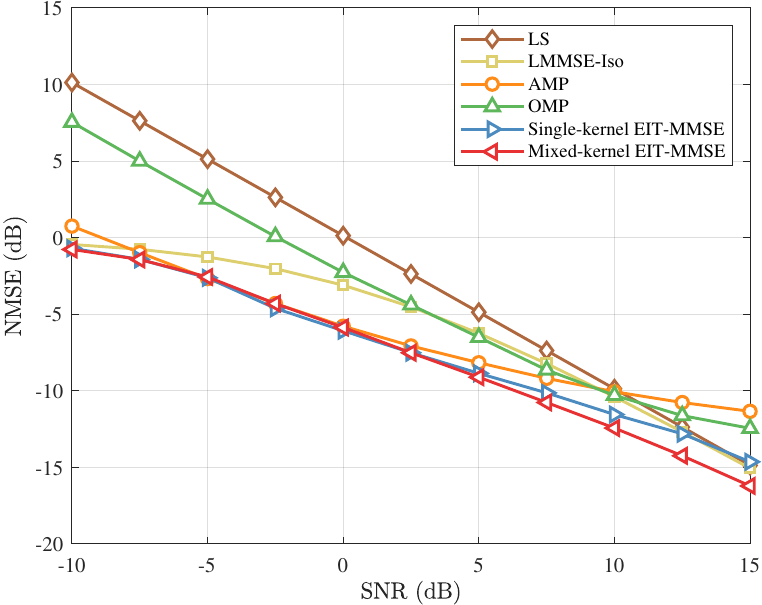}
    \caption{GPR-based EIT-MMSE channel estimation with EMCF and SV channel. }
    \label{fig:GPR-channel-estimation-SV}
\end{figure}

\begin{figure}[t]
    \centering
    \includegraphics[width=\widthRatio\linewidth]{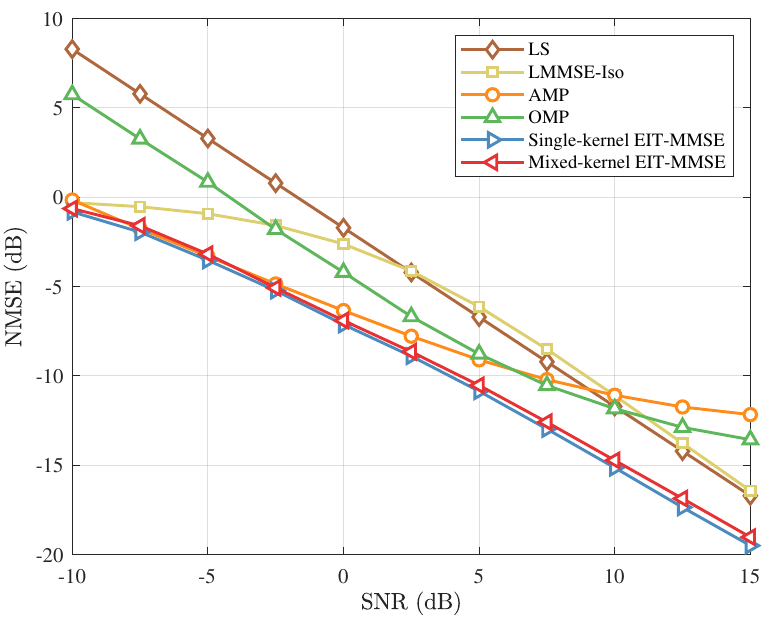}
    \caption{GPR-based EIT-MMSE channel estimation with EMCF and CDL channel. }
    \label{fig:GPR-channel-estimation-CDL}
\end{figure}

{\bf Estimator performance w.r.t. SNR.} In Fig.~\ref{fig:GPR-channel-estimation-SV}, the proposed GPR estimators are compared with various baseline channel estimators under SV channel model. Fig.~\ref{fig:GPR-channel-estimation-CDL} presents the results under the standard CDL channel model. All the estimators work in the {\it single-shot learning} regime without historical channel data. Simulation results in Fig.~\ref{fig:GPR-channel-estimation-SV} and Fig.~\ref{fig:GPR-channel-estimation-CDL} have demonstrated that, both the proposed single-kernel GPR and the multi-kernel GPR EIT-MMSE methods can outperform the baseline algorithms in terms of normalized mean square error (NMSE) across an SNR range of $-10 \sim 15\,{\rm dB}$. The mixed-kernel EIT-MMSE method outperforms all its rivals, mainly due to its strongest capability to match the EM correlation laws in the received pilots. The single-kernel EIT-MMSE method is superior to all the baseline estimators, which demonstrates that the proposed EMCF better captures the fundamental laws of the EM channel. In other words, the EMCF provides better understanding of the wireless channel compared with the sparsity understanding in the AMP and OMP algorithms, and the isotropic fading understanding in the isotropic LMMSE algorithm~\cite{demir2022channel}. Note that with both the SV channel model and the CDL channel model, the proposed EIT-MMSE methods outperform the baselines. However, the mixed-kernel EIT-MMSE method does not always achieve the best performance, which is mainly due to over-parameterization.

\section{Conclusions} \label{sec6-conclusion} 
\red{In this paper, we proposed an EIT-based statistical channel model with simplified parameterization, which is in the form of the EMCF. 
Thanks to the favorable analytic properties of the EMCF, we further proposed the EIT-Cov covariance estimator by fitting the EMCF to the observed noisy channel coefficients. The EIT-Cov covariance estimator allows single-shot learning from the observed channel coefficients, which leads to the EIT-MMSE channel estimator that integrates EM knowledge into channel estimation. 
Finally, we numerically tested the proposed EIT-Cov and EIT-MMSE estimators in uplink channel estimation with practical SV and CDL models. The EIT-Cov method outperforms the traditional sample covariance method, and the EIT-MMSE method achieves improved performance over traditional isotropic LMMSE estimators, CS-based channel estimators, and FBS-based estimators. } 

For future works, Doppler features can be extracted by tuning the EMCF to fit the time-varying channel, thus enabling the application of the EMCF to various channel inference tasks, including accurate channel prediction, fast channel tracking, and improved channel interpolation/extrapolation.  


\appendices

\section{Proof of the correlation integral (15)} \label{app:proof_correlation_integral}

We only prove the real displacement case, i.e., ${\bf r} = {\bf x} - {\bf x}'\in\mathbb{R}^3$. The complex displacement case is proved by the uniqueness of analytic continuation. 

Let the standard orthonormal basis be $\{{\bf e}_1, {\bf e}_2, {\bf e}_3\}\subset \mathbb{R}^3$ with an ``1'' in the $i$-th component of ${\bf e}_i$. Choose another orthonormal basis $\{\hat{\bf e}_1, \hat{\bf e}_2, \hat{\bf e}_3\}\subset\mathbb{R}^3$ with $\hat{\bf e}_3$ aligned to $\hat{\bf r}$. Due to its orthonormality, the matrix ${\bf R}({\bf x}; {\bf x}')\in\mathbb{C}^{3\times 3}$ is decomposed as 
\begin{equation}
    {\bf R} = \sum_{i,j=1}^{3} [\tilde{\bf R}]_{ij} \hat{\bf e}_i \hat{\bf e}_j\T, \label{eqn:base_transform}
\end{equation}
where $[\tilde{\bf R}]_{ij} := \hat{\bf e}_i\T {\bf R} \hat{\bf e}_j$. Define $\kappa_i = \hat{\bm \kappa}\T \hat{\bf e}_i$. The correlation integral can be written in its components as 
\begin{equation}
    [\tilde{\bf R}]_{ij} = \frac{\sigma^2}{8\pi}\int_{S^2} (\delta_{ij} - \kappa_i\kappa_j)e^{\ri \beta \kappa_3} {\rm d}S, 
\end{equation}
where $\beta = k_0 |{\bf r}|$. Express the above integral in spherical coordinate with $\kappa_3 = \cos\theta, \kappa_1 = \sin\theta\cos\varphi, \kappa_2 = \sin\theta\sin\varphi$, using ${\rm d}S = \sin\theta{\rm d}\theta{\rm d}\varphi$ and the fact that 
\begin{equation}
    f_n(\beta) := \int_{-1}^{1} x^n e^{\ri \beta x}{\rm d}x = \int_{0}^{\pi} \cos^n(\theta)e^{\ri \beta \cos\theta}\sin\theta{\rm d}\theta, 
\end{equation}
we obtain~\eqref{eqn:derivative_correlation_integral}. The proof is completed by applying the base transform~\eqref{eqn:base_transform} from $\{{\bf e}_i\}$ to $\{\hat{\bf e}_i\}$. 

\begin{figure*}[t]
    \centering
    \hrulefill
    \vspace*{8pt}
    \begin{equation} \label{eqn:derivative_correlation_integral}
        \begin{aligned}
        \tilde{\bf R} & = \frac{\sigma^2}{8\pi} \int_{0}^{2\pi}{\rm d}\varphi \int_{0}^{\pi}{\rm d}\theta e^{\ri\beta\cos\theta}\sin\theta \cdot \left({\bf I}_3 -\begin{bmatrix} \sin^2\theta\cos^2\varphi& \sin^2\theta\cos\varphi\sin\varphi & \sin\theta\cos\theta\cos\varphi\\ \sin^2\theta\cos\varphi\sin\varphi& \sin^2\theta\sin^2\varphi &\sin\theta\cos\theta\sin\varphi \\ \sin\theta\cos\theta\cos\varphi &\sin\theta\cos\theta\sin\varphi &\cos^2\theta  \end{bmatrix} \right) \\
        & = \frac{\sigma^2}{4} f_0{\bf I}_3 - \frac{\sigma^2}{8} \begin{bmatrix} (f_0 - f_2) & 0 & 0 \\ 0 & (f_0-f_2) & 0 \\ 0 & 0 & 2f_2 \end{bmatrix}\\
        & = \frac{\sigma^2}{8}\left((f_0 + f_2) {\bf I}_3 + (f_0 - 3f_2) {\bf e}_3 {\bf e}_3\T\right). 
        \end{aligned}
    \end{equation}
    \vspace*{8pt}
    \hrulefill
\end{figure*}

\section{Proof of \textbf{Lemma 1}} \label{app:proof-lemma-1}
Since $\sigma^2$ is only a multiplicative factor in~\eqref{eqn:Four-dimensional-STCF}, the derivative w.r.t. $\sigma^2$ is easy to obtain. Thus, following~\eqref{eqn:def_Sigma_of_w}, we focus on evaluating the derivative of ${\bf \Sigma}({\bf w})$ w.r.t. ${\bf w}$. We notice that ${\bf \Sigma}({\bf w})$ is a matrix-valued function of the complex vector variable ${\bf w}$, and more importantly, each matrix element is a multi-variable {\it analytic function} of ${\bf w}$. Following standard analytic differential techniques and using the recurrence formula\footnote{This can be verified by differentiating both sides of~\eqref{eqn:def-fn-function} with respect to $\beta$, and swapping the integral and the differential operators.} $f_n^{(k)}(\beta) = \ri^k f_{n+k}(\beta)$, we arrive at 
\begin{equation}
    \begin{aligned}
        \frac{\partial {\bf \Sigma}({\bf w})}{\partial {\bf w}(k)}&= \frac{1}{8}\left[\ri (f_1 + f_3)\hat{\bf w}(k) {\bf I}_3 + \ri(f_1-3f_3)\hat{\bf w}(k) \hat{\bf w}\hat{\bf w}\T\right. \\
        & + \left. (f_0 -3f_2) (\partial_k \hat{\bf w}\cdot \hat{\bf w}\T + \hat{\bf w} \cdot\partial_k \hat{\bf w}\T) \right],
    \end{aligned}
    \label{eqn:derivative-Sigma-wrt-w-appendix}
\end{equation}
where $\partial_k:=\partial/\partial {\bf w}(k)$, $\partial_k \hat{\bf w} = |{\bf w}|^{-1}(\hat{\bf e}_k - (\hat{\bf w}(k))\hat{\bf w})$, and all the functions $f_n(\cdot)$ are evaluated at $\beta = |{\bf w}|$. $\hat{\bf e}_k$ denotes the unit vector with the only ``1'' located at the $k$-th component. Notice that, usually in order to fully describe the derivative of a complex-valued function $g(w)$ w.r.t. a complex argument $w$, two derivatives should be involved, i.e., $\partial g/\partial w$ and $\partial g/\partial w^*$. Fortunately, due to the analyticity of ${\bf \Sigma}({\bf w})$, we have
\begin{equation}
    \frac{\partial {\bf \Sigma}}{\partial {\bf w}_k^*} = {\bf O}. 
    \label{eqn:analyticity-simplified}
\end{equation}
Combining~\eqref{eqn:derivative-Sigma-wrt-w-appendix} and~\eqref{eqn:analyticity-simplified}, the Wirtinger derivative of ${\bf K}_{\rm EM}$ w.r.t. ${\bm \mu}(k)$ can be evaluated as 
\begin{equation}
    \frac{\partial {\bf K}_{\rm EM}}{\partial {\bm \mu}(k)} = \sigma^2 \left(\frac{-\ri}{C(\mu)} \frac{\partial {\bf \Sigma}}{\partial {\bf w}(k)} - \frac{C'(\mu){\bm \mu}(k)}{C^2(\mu)\mu} {\bf \Sigma}\right),
\end{equation}
where $C'(\mu) = \mu^{-2}(\mu \cosh(\mu) - \sinh(\mu))$. This completes the proof.

\ifCLASSOPTIONcaptionsoff
 \newpage
\fi

\footnotesize

\bibliographystyle{IEEEtran}

\bibliography{IEEEabrv, bibfile}

\end{document}